\documentclass[showpacs,floatfix,twocolumn]{revtex4-1}
\usepackage{graphicx,psfrag,amsmath,amssymb,amsfonts,latexsym,color,epsf,dcolumn,graphpap}
\usepackage{subfig}
\usepackage{float}
\usepackage{lipsum}
\usepackage{enumerate}

\usepackage{tikz-cd} 
\usepackage{tikz}
\usetikzlibrary{shapes.geometric, arrows}
\usepackage{physics}

\usepackage{tikz,xcolor,hyperref}

\definecolor{lime}{HTML}{A6CE39}
\DeclareRobustCommand{\orcidicon}{
	\begin{tikzpicture}
	\draw[lime, fill=lime] (0,0) 
	circle [radius=0.16] 
	node[white] {{\fontfamily{qag}\selectfont \tiny ID}};
	\draw[white, fill=white] (-0.0625,0.095) 
	circle [radius=0.007];
	\end{tikzpicture}
	\hspace{-2mm}
}

\foreach \x in {A, ..., Z}{%
	\expandafter\xdef\csname orcid\x\endcsname{\noexpand\href{https://orcid.org/\csname orcidauthor\x\endcsname}{\noexpand\orcidicon}}
}

\definecolor{red}{rgb}{1,0,0}
\definecolor{blue}{rgb}{0,0,1}
\definecolor{skyblue}{rgb}{0,0,.5}
\definecolor{green}{rgb}{0,1,0}
\definecolor{orange}{cmyk}{0,.4,1,0}

\begin{document}
\title{Photon generation and entanglement in a double superconducting cavity}

\author{Cruz I.~Velasco$^1$\orcidA}
\author{Nicol\'as F.~Del Grosso$^{1,2}$\orcidN}
\author{Fernando C. Lombardo$^{1,2}$\orcidF}
\author{Alejandro Soba$^3$ }
\author{Paula I.~Villar$^{1,2}$ }

\affiliation{$^1$ Departamento de F\'\i sica {\it Juan Jos\'e
 Giambiagi}, FCEyN UBA, Facultad de Ciencias Exactas y Naturales,
 Ciudad Universitaria, Pabell\' on I, 1428 Buenos Aires, Argentina }
 \affiliation{$^2$ IFIBA CONICET-UBA, Facultad de Ciencias Exactas y Naturales,
 Ciudad Universitaria, Pabell\' on I, 1428 Buenos Aires, Argentina }
\affiliation{$^3$ Centro At\'omico Constituyentes, Comisi\'on Nacional de Energ\'\i a At\'omica, Avenida General Paz 1499, San Mart\'\i n, Argentina}
\begin{abstract}
We study the dynamical Casimir effect in a double superconducting cavity in a circuit quantum electrodynamics architecture. Parameters in the quantum circuit are chosen in such a way the superconducting cavity can mimic a double cavity, formed by two perfectly conducting outer walls and a dielectric one, with arbitrary permittivity separating both halves. We undertake a spectral analysis of the cavity, showing that the spectrum varies significantly depending on the values of the susceptibility of the dielectric mirror and the relative lengths of both cavities. We study the creation of photons when the walls oscillate harmonically with a small amplitude. Furthermore, we explore  the possibility of entangling two uncoupled cavities, starting from a symmetric double cavity and having both of its halves become uncoupled at a later given instant. We consider both cases: (i) when the field is initially in a vacuum state and (ii) the situation in which photon creation via the dynamical Casimir effect has already taken place. We show that the cavities become entangled in both cases but, in the latter, the quantum correlation between individual modes can be greatly increased at the cost of diminishing the entanglement between most pairs of modes.

\end{abstract}

%\author{N.F.~Del Grosso, F.C.~Lombardo, P.I.~Villar, %F.D.~Mazzitelli}
%\affiliation{$^1$ Departamento de F\'\i sica {\it Juan Jos\'e Giambiagi}, FCEyN UBA and IFIBA CONICET-UBA, Facultad de Ciencias Exactas y Naturales, Ciudad Universitaria, Pabell\' on I, 1428 Buenos Aires, Argentina.\\}
\date{today}
\maketitle
%====================================================================
%\begin{abstract}
\noindent 

%\end{abstract}
%\pacs{03.70.+k, 12.20.Ds, 42.50.Lc, 42.65.Yj}
\maketitle
%\pdfoutput=1
%====================================================================
\section{Introduction}\label{sec:intro}

The dynamical Casimir effect (DCE), in which the mechanical oscillation of the position of a mirror inside a cavity produces particles from an initial vacuum state \cite{reviews, moore, DeWitt, FullingDavies, Dodonov, dalvit shaker, Plunien, crocceScalar}, remains one of the most surprising predictions of quantum field theory. However, the magnitude of the frequencies at which mechanical oscillations must occur to produce an appreciable number of photons, can not be reached with current technology for massive mirrors \cite{fiftyYears}. Consequently no observations of this phenomena has taken place to this date.

To bypass this complication, a lot of systems have been studied as to explore the possibility of varying the boundary conditions of a field inside the cavity, without the mechanical movement of the walls. One of the more promising systems where this can be achieved are quantum superconducting circuits, which are already  being used in quantum information and quantum computing widely \cite{qubit_circuit, computacionChina, computacionCanada}. In such a case,  an external and time dependent magnetic field is used to vary the magnetic flux through a superconducting quantum interference device (SQUID), and through it, change the effective length of a superconducting waveguide, simulating a moving mirror. This type of systems has led to the observation of this effect \cite{squidDemo}. Even so, experiments have yet to show the conversion of mechanical energy (associated with the movement of mirrors) into photons. 

In this work we present a study of a circuit conformed by two waveguides that are coupled between them by a SQUID and each waveguide has  also  another SQUID at its other end. This system, under a certain choice of parameters, can easily be linked to one composed by a one dimensional double cavity that has perfect conductors as its external walls and a semitransparent dielectric wall between them. The two systems share similar generalized boundary conditions \cite{contornosGenerales}, which admits an equivalent mathematical description of both. However, the cavity allows for a physically more intuitive picture. Another interesting property of these systems is that, through the adjustable parameters they have, one could, in principle obtain different eigenfrequency spectrum structures. This is of great importance when studying the DCE, as different regimes of said structure can result in different results for the photon production rate \cite{crocceScalar}.

Finally, we have mentioned that superconducting circuits are relevant for quantum information. The same can be said for double cavity systems. It is a well known fact that the quantum vacuum presents entanglement between spatially separated regions \cite{entvacio}. In reference \cite{HalfEmpty}, authors studied the possibility of harvesting entanglement from a vacuum state in a simple cavity into a two cavity system by quickly introducing a mirror. This might result in the generation of two entangled cavities that could be spatially separated and used in quantum communication applications. The double cavity system treated in the present manuscript presents a natural framework to continue the study of this problem. It is also well known that the DCE can allow for the creation of entangled pair of photons \cite{dodonovEntanglement}, and even the redistribution of entanglement between field modes that already present some degree of quantum correlation \cite{entNico}. This means that, in principle, the DCE could be used to improve the results of entanglement harvesting from an initial vacuum state. 

The article is organized as follows. In Sec. \ref{sec: model} we present the model that will be used for the rest of the work.  We show how the mathematical description of the two systems is equivalent, under certain choice of parameters set. In Sec. \ref{sec: spectral}, we present an analysis of the eigenfrequency spectrum of the double cavity for different choices of the electric susceptibility of the dielectric wall and the positions of the three walls. In Sec. \ref{sec: creation}, we show the results for photon creation as a consequence of the harmonic oscillation of the perfectly conducting walls of the double cavity. We carry out the study both through analytical and numerical means. For the analytical results, we employ a method known as multiple scale analysis. The numerical method consists of an integration of the exact differential equations that govern the state of the field with time dependent boundary conditions. Further, in Sec. \ref{sec: entanglement}, we study the possibility of using the DCE to increase the entanglement between cavities that have been decoupled with respect to the previously studied case of an initial vacuum state. Finally, in Sec. \ref{sec: conc} we present the conclusions of the work.

\section{The Model} \label{sec: model}

\begin{figure}
    \centering
    \includegraphics[width = 8cm]{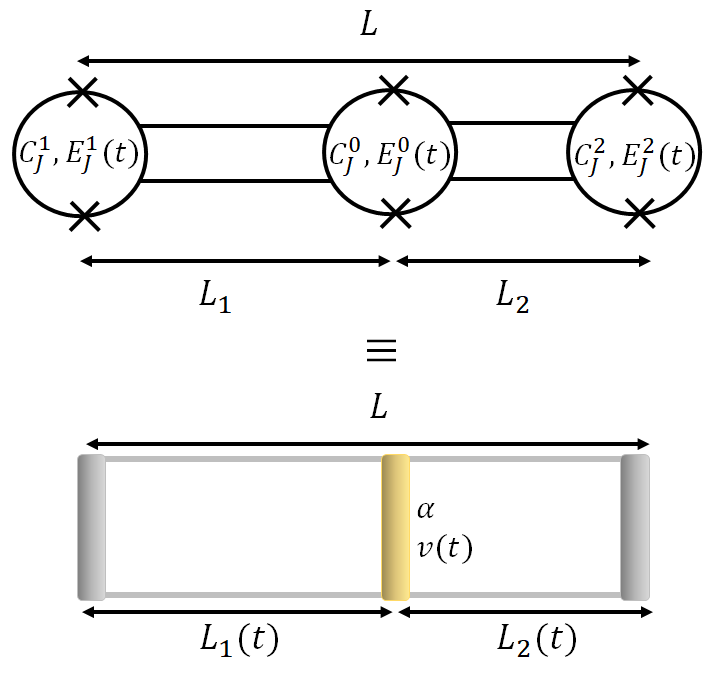}
    \caption{Schematics of the superconducting quantum circuit and its double cavity analogue.}
    \label{fig: esquema}
\end{figure}

The system is a superconducting circuit of length $L$ schematized in the upper half of Fig. \ref{fig: esquema}. It is composed of two one-dimensional superconducting waveguides of length $L_1=(L+L_0)/2$ (left) and $L_2=(L-L_0)/2$ (right) which are coupled by a SQUID in $x = 0$. Both waveguides also have SQUIDs in their other ends, located at $x = -L_1$ (in the left's case) and $x = L_2$ (in the right's).  We consider both of the waveguides to be  characterized by the same capacitance and inductance per unit length, $c_a$ and $l$ respectively. In such a situation, it is convenient to describe the state of the electromagnetic field inside the circuit in terms of the \textit{phase field}: $\varphi(x,t) = \int^t dt' E(x,t')$. The Lagrangian density for the system can be written as \cite{squidPauyFer}
\begin{equation}
    \mathcal{L}_{\rm circuit}=\frac{1}{2}\Big(\frac{1}{2e}\Big)^2(C(\partial_t\varphi)^2-v_{w}^2c_a(\partial_x\varphi)^2)-E(t)\varphi^2
    \label{eq: lagrangianCirc}
\end{equation}
where we have defined $v_w = (lc_a)^{-1/2}$,
\begin{equation}
    C = (c+2\delta(x)C^2_J+2\delta(x+L_1)C_J^1+2\delta(x-L_2)C_J^2)
\end{equation}
and
\begin{align}
    E(t)=\delta(x)E_J^0 \cos(f_0(t))+\delta(x+L_1)E_J^1 \cos(f_1(t))\nonumber\\+\delta(x-L_2)E_J^{2} \cos(f_2(t)).
\end{align}

Here $C^0_J$ and $E^0_J$ are, respectively, the Josephson capacitance and energy of the central SQUID. The SQUIDS at the outer ends of the circuit are also characterized by their capacitance ($C_J^{1,2}$) and energy ($E_J^{1,2}$). The Josephson energy of all three SQUIDs can be manipulated by modifying the magnetic flux that goes through it. The variation of said magnetic flux is characterized by $f_i(t)$ ($i = 0,1,2$). In the previous equations and for the rest of this work we consider $\hbar = c = 1$.

The field equation for $\varphi$ can be readily obtained from Eq. (\ref{eq: lagrangianCirc}) and reads
\begin{equation}
    \square\varphi(x,t) = 0.
    \label{eq: FieldEq}
\end{equation}
The field must be continuous and satisfy boundary conditions imposed by the presence of the three SQUIDs. These conditions can be obtained from Eq. (\ref{eq: lagrangianCirc}) and consist of discontinuities of the spatial derivative of $\varphi$, in $x = -L_1$, $x = L_2$ and $x = 0$. Particularly, for $x=0$ we get
\begin{align}
    \partial_x\varphi(t,0^+)-\partial_x\varphi(t,0^-)=C_J^{0\prime}\partial_t^2\varphi(t,0)\nonumber\\+E_J^{0\prime}(t)\varphi(t,0).
    \label{eq: boundaryCircuit}
\end{align}
Herein, we have defined $C_J^{0\prime} = C_J^0/v_w^2c_a$ and $E_J^{0\prime}(t) = (2e)^2(2E^0_J/v_w^2c_a)\cos(f_0(t))$.

The other two conditions are analogous to  Eq. (\ref{eq: boundaryCircuit}) and can be used to decouple the field inside the circuit from any external system by choosing $E_J^{1}=E_J^{2}=E_J$ with $e^2 E_J/v_w^2 c_a\gg 1$ \cite{squidPauyFer}. This will allow us to consider that $\varphi(x,t) = 0$ for $x<-L_1$ and $x>L_2$. For simplicity, we also consider $C_J^{1,2} = 0$ and that both SQUIDs are subjected to a time dependent magnetic flux. Thus, the boundary condition simplifies, allowing us to assume
\begin{align}
    0=\varphi(\mp L_{1,2})+\frac{1}{E_J^{\prime}(t)}\partial_x\varphi(\mp L_{1,2}) \approx\nonumber\\\approx \varphi\Big(\frac{\mp L-L_0}{2} + \frac{1}{E_J^{\prime}(t)}\Big). \label{eq: CircDirichlet}
\end{align}
We have defined $E_J^{\prime}(t) = (2e)^2\frac{2E_J}{v_w^2c_a}\cos(f(t))$ and $f(t) = f_1(t) = f_2(t)$. It can be noted that the boundary conditions introduced in (\ref{eq: CircDirichlet}) can be interpreted as in-phase changes in the effective lengths of the two waveguides which can be characterized by
\begin{equation}
    \Delta L(t) = L_0-\frac{2}{E_J^{\prime}(t)}.
    \label{eq: dispCirc}
\end{equation}
This fact can be seen as a consequence of the mathematical analogy between superconducting circuits and cavities. For the particular choice of parameters made for our circuit, an analogous system can be found in a one-dimensional double cavity of length $L = L_1 + L_2$ (as schematized in Fig. \ref{fig: esquema}), with perfectly conducting outer walls and a semi-transparent dielectric wall separating both halves in $x = 0$. This wall is considered to be infinitesimally thin, which is a reasonable approximation if we consider this type of systems \cite{membrana_fina}. This configuration results in a left cavity of length $L_1$ and a right cavity of length $L_2$, both sharing the central dielectric wall. The dielectric membrane is characterized by its electric permittivity $\epsilon$. A time dependent electric potential $V(t)$ is also applied on this wall. 

Further, we shall consider a scalar field $\hat{\phi}$ inside of the cavity (as a simplified model of the electromagnetic field). It can be described by the following Lagrangian density \cite{contornosGenerales}
\begin{equation}
    \mathcal{L} = \frac{1}{2}\Big(\varepsilon(\partial_t \hat{\phi})^2-(\partial_x \hat{\phi})^2-V(t)\hat{\phi}^2\Big)
    \label{eq: lagrangianCav}
\end{equation}
where
\begin{equation}
    \varepsilon = 1+\delta(x)\alpha
\end{equation}
and
\begin{equation}
    V(t)=\delta(x)v(t).
\end{equation}
From now on, we will use $\alpha$ and $v(t)$ to characterize the dielectric membrane. For simplicity, we will refer to $\alpha$ as the wall's electric susceptibility, and $v(t)$ as the electric potential applied to it, despite the difference in units of both parameters with said quantities.

Herein,  we have taken a coordinate system with its origin in the position of the dielectric wall, which results in the position of the perfectly conducting walls being set at $x = -L_1$ for the left one, and $x = L_2$ for the one in the right. 

We shall consider the case where the position of the perfectly conducting walls may vary in time ($L_1(t)$, $L_2(t)$) while the dielectric wall stays static in $x = 0$. Particularly, we assume that the movement of the external walls is such that the total length of the double cavity does not vary, i.e. $L_1(t)+L_2(t) = L$ remains a constant. Then, it is convenient to characterize the movement of the walls through a displacement parameter $\Delta L(t)$ such that,
\begin{equation}
    \Delta L(t) = L_1(t)-L_2(t).
\end{equation} 
This definition also allows us to write $L_{1,2} = (L\pm \Delta L)/2$. $\Delta L(t)$ defined here is completely analogous to the one defined in Eq. (\ref{eq: dispCirc}). It is important to note that this equivalence  means the moving walls can be simulated by a variation of the magnetic flux through the SQUIDs at the extremes of the circuit.

From Eq. (\ref{eq: lagrangianCav}) we can derive the field equation, which is equivalent to that obtained for the circuit (Eq. (\ref{eq: FieldEq})). Similarly,  the field $\hat{\phi}$ must also be continuous for all values of $x$ along the double cavity. The boundary conditions appear as a consequence of the presence of the external perfectly conducting mirrors and the central dielectric wall. For the former, we assume Dirichlet conditions in $x=-L_1(t)$ and $x = L_2(t)$. For the latter, located in $x = 0$ we ask for the continuity of the field and, a discontinuity of the spatial derivative of the field, obtained from Eq. (\ref{eq: lagrangianCav}), which reads
\begin{align}
    \partial_x\hat{\phi}(t,x=0^+)-\partial_{x}\hat{\phi}(t,x=0^-)=\alpha\partial_t^2\hat{\phi}(t,0)\nonumber\\+v(t)\hat{\phi}(t,0)).
    \label{eq: boundaryDielectric}
\end{align}
By comparing Eq. (\ref{eq: boundaryCircuit}) with (\ref{eq: boundaryDielectric}), we obtain a direct analogy between the susceptibility $\alpha$ of the dielectric wall with the capacitance of the central SQUID $C_J^{0\prime}$, and the potential applied to the membrane $v(t)$ with the Josephson energy of that same SQUID $E_J^{0\prime}(t)$.

As we have shown the equivalence between both systems, in the following we shall refer to the  mechanical cavity as it provides a more intuitive picture.

\subsection{Eigenfunctions}

Let us now consider the case where the double-cavity is in a static regime, meaning neither $\Delta L$ nor $v$ vary in time. In this case, all of the boundary conditions for the field inside the cavity become independent of time. In this case, we can find a basis of eigenfunctions that are solutions to the field equation (Eq. (\ref{eq: FieldEq}))  and satisfy the boundary conditions. These functions are given by \cite{hasan}
\begin{align}
    u_m=\frac{e^{-i\omega_m t}}{\sqrt{2\omega_m}N_m}\Big[\theta(-x)\sin(k_m L_2)\sin(k_m (x+L_1)) \nonumber\\- \theta(x) \sin(k_m L_1)\sin(k_m (x-L_2)) \Big],
    \label{eq: NormalModes}
\end{align}
where $k_m$ corresponds to the eigenfunction's wave-number, $\omega_m$ to its temporal frequency (which in the one dimensional case is equal to $k_m$) and $N_m$ is a normalization factor. The functions form an orthogonal basis if the following generalized Klein-Gordon inner product is satisfied \cite{hasan}
\begin{align}
    (f,g)_{KG} = i\int_{-L_1}^{L_2}(\dot{f}(t,x)g^*(t,x)\nonumber\\-f(t,x)\dot{g}^{*}(t,x))(1+\delta(x)\alpha)dx.
\end{align}
The normalization factor $N_m$ is defined in such a way that the basis is orthonormal ($(u_n,u_l)_{KG} = \delta_{nl}$) and reads
\begin{align}
    N_m^2 = \frac{1}{2} \Big( L_1 \sin^2(k_m L_2) + L_2 \sin^2(k_m L_1) \nonumber\\+ \frac{1}{k_m} \sin(k_m L)\sin(k_m L_1) \sin(k_m L_2)\Big). 
\end{align}
The values of $k_m$ are such that the eigenfunctions are consistent with Eq. (\ref{eq: boundaryDielectric}). This condition implies that 
\begin{equation}
    \cos(k_m\Delta L)-\cos(k_m L)=\frac{2\sin(k_m L)}{k_m \Big(\alpha+\frac{v}{k_m^2}\Big)},
    \label{eq: trascendente}
\end{equation}
which defines the admissible values of $k_m$. This transcendental equation will be studied further in Section \ref{sec: spectral}.

By using these eigenfunctions we can expand the field operator as \cite{Grenier}
\begin{equation}
    \hat{\phi}(t,x) = \sum_m (\hat{a}_m u_m(t,x) + \hat{a}^{\dagger}_m u_m^{*}(t,x)),
\end{equation}
where $\hat{a}_m$ are the bosonic operators corresponding to the different photon modes.

If we then consider moving boundaries or a time varying potential applied to the central wall, the functions $u_m(t,x)$ can be expressed in terms of an instantaneous basis \cite{Dodonov, baseinst}
\begin{equation}
    u_m(t,x) = \sum_n Q_n^m(t) \Phi_n(x,t)
\end{equation}
where $Q_n^m(t)$ are time dependent coefficients and

\begin{widetext}
\begin{equation}
    \Phi_n(x,t) = \frac{1}{N_n}\Big[\theta(-x)\sin(k_n(t) L_2(t))\sin(k_n(t) (x+L_1(t))) -\theta(x) \sin(k_n(t) L_1(t))\sin(k_n(t) (x-L_2(t))) \Big].
    \label{eq: baseInstantanea}
\end{equation}
\end{widetext}
It is important to note that the wave-numbers dependence on time comes from the time dependence of $\Delta L$ and $v$ in Eq. (\ref{eq: trascendente}). The functions given in (\ref{eq: baseInstantanea}) form an orthonormal basis corresponding to the generalized inner product
\begin{align}
    (f,g) = \int_{-L_1}^{L_2} f(x)g(x)(1+\delta(x)\alpha)dx,
\end{align}
that we use throughout the work.

\section{Double cavity spectrum} \label{sec: spectral}

In order to make a complete description of the double cavity, we need to study the solutions to Eq. (\ref{eq: trascendente}). For simplicity, we consider the case where $v(t) = 0$, i.e. there is no potential applied at the dielectric wall, and thus, the solutions will only depend on $L$, $\Delta L$ and $\alpha$. We assume the value of $L$ to be fixed, and use it to define the dimensionless quantities $\Delta L/L$ and $\alpha/L$. Additionally, we consider that the walls of the cavity are not moving, which implies that $\Delta L$ is constant. We must note that as Eq. (\ref{eq: trascendente}) is a transcendental equation, it cannot be solved analytically for general values of $\Delta L/L$ and $\alpha/L$. There are however, two sets of values of $\alpha/L$ where approximate analytical solutions can be found, namely $k_m\alpha \ll 1$ and $k_m\alpha \gg 1$. Furthermore, in the limit cases $\alpha/L = 0$ and $\alpha/L \rightarrow \infty$, the analytical solutions become exact. 

We note now that in general, we must consider the value of $k_m \alpha$ instead of just $\alpha/L$ when approximating because of the way $\alpha$ appears in Eq. (\ref{eq: trascendente}), multiplied by $k_m$. This caveat is particularly important when dealing with lower values of $\alpha$, as we will see later. 

We shall start with the general case of the dielectric wall, and show later the cases where analytical solutions can be applied.

\subsection{Dielectric wall}

\begin{figure}[th!]
    %\centering
    \includegraphics[width = 8.5cm]{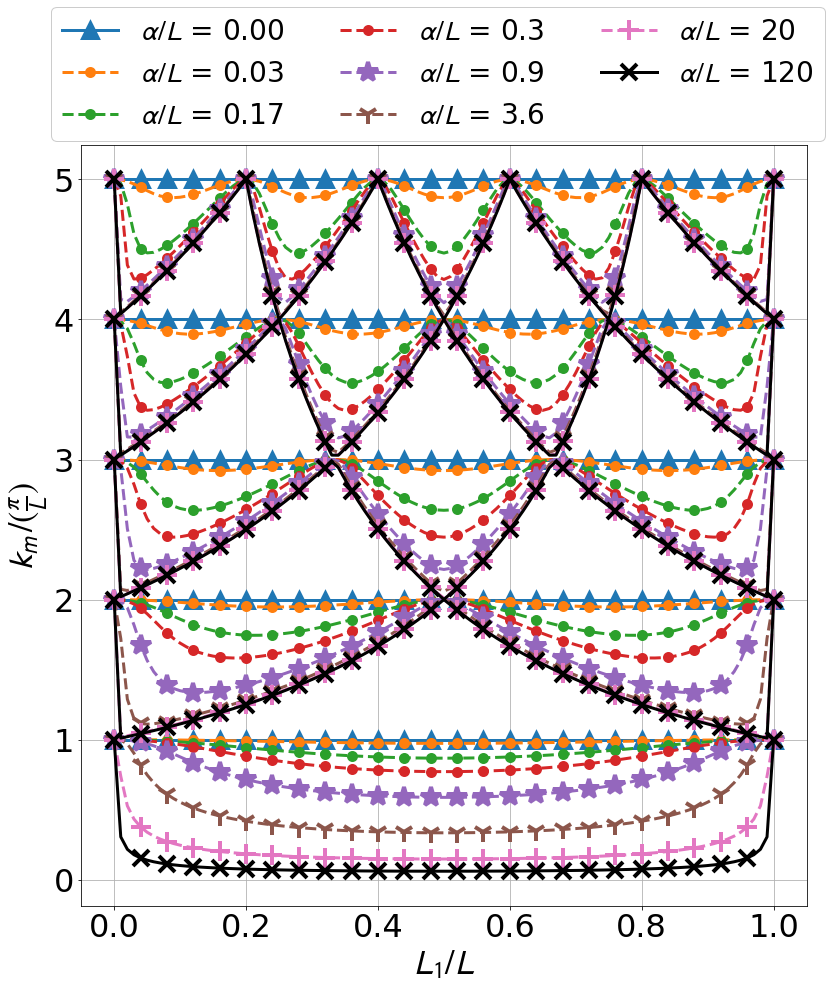}
    \caption{Wave-number spectrum as a function of the length of the left cavity ($L_1/L$) for different values of $\alpha/L$ between 0 and 120. The figure shows said values for the first 5 modes ($k_0$,$k_1$,$k_2$,$k_3$,$k_4$).}
    \label{fig: Spectrum}
\end{figure}

For cases in which the susceptibility is such that $k_m\alpha \sim 1$, we cannot use neither of both approximations and so we must solve the Eq. (\ref{eq: trascendente}) numerically. To do this we use a similar method to the one employed in \cite{squidPauyFer}. In Fig. \ref{fig: Spectrum} we show the first five wave numbers, corresponding to $k_0$, $k_1$, $k_2$, $k_3$, and $k_4$ for different lengths of the left cavity ($L_1/L$). All of the wave numbers are also plotted for different values of the electric susceptibility ($\alpha/L$) according to the colors detailed in the legend.
In Fig. \ref{fig: Spectrum}, we can see that for $\alpha/L = 0$  the wave-numbers are given by straight horizontal lines which are integer multiples of $\pi/L$ and thus,  for this case $k_m$ results independent of the relative size of the left cavity, and in extension of $\Delta L/L$. This is consistent with the fact that for $\alpha/L  = 0$ the dielectric wall becomes completely transparent, and thus, the wavenumbers are those corresponding to the case of a single cavity of length $L$. On the other hand, for high values of the susceptibility ($\alpha/L = 120$), we can note that the wave-numbers (except $k_0$) are proportional to either $L_1^{-1}$ or $(L-L_1)^{-1} = L_2^{-1}$. In this case, the value of $\alpha/L$ turns the dielectric wall into an almost perfectly conducting wall. This results in the system becoming one of two cavities with perfectly conducting walls at their ends one of which they share among each other. As a consequence of this, the wavenumbers (with $m>1$) become either integer multiples of $\pi/L_1$ or $\pi/L_2$. In the particular case of the first mode ($k_0$), we see that as the value of $\alpha/L$ increases, its value decreases. As we will see in Section \ref{sec: perfectcond}, this mode becomes irrelevant as the susceptibility becomes larger.

For intermediate values of $\alpha/L$ we get a continuous connection between the two limit cases where, as the value of $\alpha/L$ increases (starting from $\alpha/L = 0$), the wave-numbers gain a dependence on $L_1$ and, consequently, on  $\Delta L$. As the value of the susceptibility rises, the dependence becomes more apparent, until we get $k_m \propto L_{1,2}^{-1}$. Another feature is that the more energetic modes coincide with the latter extreme value ($k_m \propto L_{1,2}^{-1}$) for lower values of $\alpha/L$ ($\alpha/L = 0.9$ for $k_4$) than the ones that are less energetic ($\alpha/L = 3.6$ for $k_1$). Finally, the ``zeroth"-mode's wave-number, tends to vanish at a slower rate than for which the rest obtain their extreme value. This can be seen, for example, in the fact that while for $\alpha/L = 3.6$ all of the higher modes have already taken their $\propto L_{1,2}^{-1}$ shape, $k_0$ has only decreased to approximately a third of its value at $\alpha/L = 0$.

\begin{figure}[th!]
    \centering
    \includegraphics[width = 8cm]{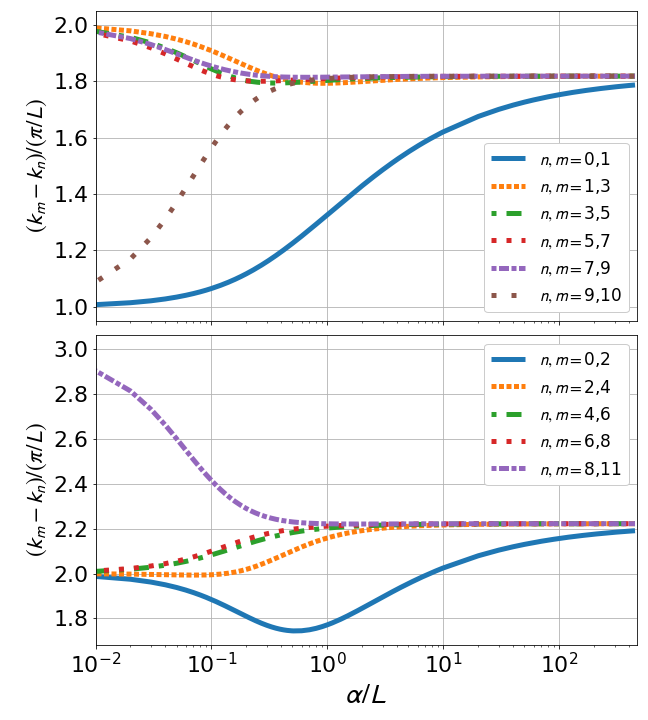}
    \caption{Difference between wave-numbers as a function of $\alpha/L$ for $L_1/L = 0.6$. The differences considered are those between wave-numbers corresponding to eigenfunctions that, in the case of $\alpha/L\rightarrow\infty$, localize either in the left cavity (up) and in the right cavity (down).}
    \label{fig: Distance}
\end{figure}

An important factor to consider, mostly for the study of particle creation via the DCE, is that of the spacing between modes, as an equidistant spectrum (i.e. $k_{m+1}-k_m$ is constant for a given set of $m$) leads to very different results than for one that is non-equidistant \cite{crocceScalar}. From the previous results, we can see that for both limit cases, we get equidistant spectra. In the case of $\alpha = 0$ we get $k_{m+1}-k_m = k_0$ for all $m$, while for $\alpha/L \rightarrow \infty$ two different equidistant spectra $k_{m+1}^{\pm}-k_m^{\pm} = k_1^{\pm}$. For more general values of $\alpha$ we again use the numerical solutions to Eq. (\ref{eq: trascendente}). Fig. \ref{fig: Distance} shows the difference between pairs of wave-numbers that, for large enough values of $\alpha/L$, correspond to localized modes in either the right or left cavity. The values shown in Fig. \ref{fig: Distance} correspond to the case where $L_1/L = 0.6$, the spacing between the modes will vary with the value of $L_1/L$.

We can see that the behavior both for low and high values of $\alpha/L$ are what we would expect. In the first case, we tend to have $k_m-k_n = (m-n)\pi/L$, while in the latter, we have the same value for the difference of all subsequent modes corresponding to the same cavity. For values of $\alpha/L \sim 10$ we get equidistance for the spectra corresponding to each of the cavities, however, for values of this order, the mode corresponding to $m = 0$ is still far from fulfilling this condition. This is another consequence of $k_0$ reaching its limit value of zero for values of $\alpha/L$ much higher than the rest of the modes. Anyway, for the rest of the modes, we can consider that the spectrum inside of each of the perfectly conducting cavities is equidistant for values of $\alpha/L$ of the order of 10.

In the following, we  shall discuss the  approximate solutions that can be obtained for $k_m\alpha \ll 1$ and $k_m \alpha \gg 1$. These solutions lead to a  more accurate explanation of the behavior of $k_m$ as a function of $\alpha$ and $\Delta L$ related to what we have been discussing in this section.

\subsection{Transparent wall}

We shall start with $k_m\alpha \ll 1$. We consider the dielectric wall is almost transparent, with $\alpha/L = 0$ corresponding to a fully transparent wall, that corresponds to a system consisting of a single cavity. The particular case of the transparent wall, can be represented by the equations $\sin(k_m L) = 0$, with $k_m = (m+1)\pi/L$ as the  solutions. As we have already mentioned, in this case the wave-numbers become independent of $\Delta L$ and thus,  independent of the difference between the individual cavity sizes ($L_1$ and $L_2$). This is consistent with the fact that as the wall becomes fully transparent, we are left with a single cavity instead of two. Furthermore, by introducing $k_m = (m+1)\pi/L$ in Eq. (\ref{eq: NormalModes}) we obtain
\begin{equation}
    u_m^{0}(t,x)=\frac{e^{-i\omega_m^0 t}}{\sqrt{2\omega_m^0}}\sqrt{\frac{2}{L}} \sin\bigg(\frac{(m+1)\pi}{L}\,x\bigg),
    \label{eq: modos alfa0}
\end{equation}
where $\omega_m^0 = (m+1)\pi/L$. These eigenfunctions are similar to the ones corresponding to a one-dimensional single cavity with perfectly conducting walls. Here, the first spatial derivative of $u_m^0$ becomes continuous in $x=0$, as that point looses its physical importance when the wall is transparent.

For the more general case, we can consider $k_m = (m\pi/L)(1+\kappa^0_m)$ with $\kappa^0_m\ll 1$ and $(m\pi/L)\alpha \ll 1$ and solve Eq. (\ref{eq: trascendente}) for $\kappa^0_m$ to the first order. Hence, we obtain 
\begin{eqnarray}
    k_m = \frac{(m+1)\pi}{L}\bigg(1-\frac{\alpha}{2L}\Big(1+(-1)^m\nonumber\\\times\cos\Big(\frac{(m+1)\pi}{L}\Delta L\Big)\Big)\bigg).
    \label{eq: kmTransparent}
\end{eqnarray}
As we can see in the expression above, the first order term is proportional to $(m+1)\alpha/L$ which grows as we consider higher wave numbers (higher values of $m$). This means that, for non-zero values of $\alpha$, this approximation will be worse for modes that are more energetic. We can see that the behavior described by Eq. \ref{eq: kmTransparent} is also similar to the case $\alpha/L = 0.03$ for all modes, where we essentially get oscillations with respect to the horizontal line of amplitude $\propto\alpha/L$ and a frequency that depends on the mode we are considering ($\propto (m+1)$).

\subsection{Perfectly conducting wall} \label{sec: perfectcond}

In the opposite case, $k_m\alpha \gg 1$,  we assume a very reflective dielectric wall, which  may be  considered to be an 'imperfect' conductor (a mirror). Particularly, the limit case of $\alpha/L \rightarrow \infty$ corresponds to a perfectly conducting wall separating both cavities. In this case, we obtain a system consisting of two perfectly conducting cavities that share one of their walls. The transcendental equation here simplifies to $\cos(k_m \Delta L)-\cos(k_m L) = 0$. This equation has ``three" sets of solutions: the first two solutions of the form $k_m^{\pm} = 2m\pi/(L\pm\Delta L)$ and the other one, a trivial solution $k_0 = 0$. By introducing $k_m^{\pm}$ in the eigenmodes of Eq. (\ref{eq: NormalModes}), we get
\begin{eqnarray}
    u_m^\pm(t,x) &=& \frac{e^{-i \omega^\pm_m t}}{\sqrt{2\omega_m^{\pm}}}\sqrt{\frac{4}{L\pm\Delta L}}\nonumber\\ &\times & \theta(\mp x)\sin\Big(\frac{2m\pi}{L\pm\Delta L}\Big(x\pm\frac{L\pm\Delta L}{2}\Big)\Big)
    \label{eq: LocalizedModes}
\end{eqnarray}
where $\omega^\pm=k_m^{\pm}$. Here we can see that depending on which subset of solutions we choose, the eigenfunctions become localized in either the left cavity ($x<0$) or the right one ($x>0$). This result will be important when studying entanglement in Section \ref{sec: entanglement}.

When considering $(2m\pi/(L\pm\Delta L))\alpha\gg 1$, we can solve the transcendental equation with $k_m^{\pm} = (2m\pi/(L\pm \Delta L))(1+\kappa_m^{\pm})$ for $\kappa^{\pm} \ll 1$ and  we find the following solution up to first order
\begin{equation}
    k_m^{\pm}=\frac{2m\pi}{(L\pm\Delta L)}\Big(1\mp\frac{(L\pm\Delta L)}{2\alpha(m\pi)^2}\Big).
    \label{eq: kmLoc}
\end{equation}
In this case, the approximations for the more energetic modes will be better than for those with lower $m$. This can be seen in Eq. (\ref{eq: kmLoc}) where the first order term is proportional to $L/(m\alpha)$.

As for the solution $k_0 = 0$, we assume $k_0 = (\pi/L)\kappa_0$ with $\kappa_0\ll 1$  (instead of $\alpha/L\gg 1$), and hence obtain a non-trivial solution
\begin{equation}
    k_0 = \sqrt{\frac{4L}{\alpha(L^2-\Delta L^2)}}.
    \label{eq: modo0}
\end{equation}
We note that this ``zeroth" mode wave-number depends differently on $L/\alpha$ in comparison to the other modes. Herein,  we do not have a linear dependence like for $k_m^{\pm}$, but  $\sqrt{L/\alpha}$ instead. Not only does this mean that the approximation will be valid for higher values of $\alpha/L$ than for the rest of the modes, but also that it takes on its limiting value ($k_0 = 0$) for higher values of said parameter. It is important to note that the eigenfunction corresponding to this wave-number does not localize in either cavity when the susceptibility increases. Particularly, it decreases in amplitude, vanishing as $\alpha/L \rightarrow \infty$. The fact that this mode will not become localized, while also {\textcolor{red} reaching its limit value at a slower pace}, indicates that, although all eigenfunctions corresponding to $m\geq 1$ are localized, there still is a non-localized eigenfunction for $m = 0$, that may allow us to ``couple" the two cavities. As we shall see in a later section, this feature leads to interesting results both for particle creation and entanglement.

\section{Particle creation}\label{sec: creation}

Previously, we have made a thorough study of the spectrum of the field inside the double cavity for different values of the significant parameters. We have further studied  the behavior of the field's  eigenfunctions. Hence, in this section, we can analyze the particle creation process. The results of the previous section will greatly impact on the creation rate. We shall show the benefit of having such a system in order to get, for example, the exponential growth of the number of photons in a one dimensional cavity (parametric resonance), a result that is not usually obtained in this type of systems \cite{crocceScalar,crocceVect}. We shall also analyze the particular behavior of the zeroth mode, obtaining interesting solutions that can not be obtained in a double cavity with three perfectly conducing walls. 

We consider the situation where the double cavity is initially at rest, with the difference between cavity sizes given by $\Delta L(t<0) = L_0$. At a given instant $t=0$ the walls of the cavity oscillate, following a trajectory such that the distance among the walls is kept fixed. The walls  oscillate for a period of time $t_f$, at a constant frequency $\Omega$ and an amplitude $L\epsilon$, modeled by
\begin{equation}
    \Delta L(t) = L_0 + L\epsilon \sin(\Omega t).
    \label{eq: trayectoria}
\end{equation}
For times larger than $t=t_f$, the walls stop moving and the cavity becomes static again. For simplicity, we consider that there is no potential difference being applied over the dielectric wall ($v(t) = 0$ for all $t$).

For $t<0$, the cavity can be described in terms of the static basis $u_m(t,x)$ defined in Eq. (\ref{eq: NormalModes}) and bosonic operators that correspond to photon modes in that region of the space-time, $a_m^{in}$. This reads
\begin{equation}
    \hat{\phi} = \sum_m (u_m(t,x)\hat{a}^{in\,}_m+u_m^{*}(t,x)\hat{a}_m^{in\,\dagger}).
    \label{eq: campoIn}
\end{equation}
We consider the field to be in a state $\ket{in}$, which has a well-defined number of photons of well-defined wave-number $k_m$ in $t<0$ ($\langle\hat{N}_m\rangle = \expval{\hat{a}_m^{in\,\dagger}\hat{a}_m^{in}} = N_m^0$).

As the walls move, the original basis gets continually deformed into a new one satisfying the boundary conditions $u_m(t,x)\rightarrow v_l(t,x)$. The field can be expanded in this new basis
\begin{equation}
    \hat{\phi} = \sum_l (v_l(t,x)\hat{a}^{out\,}_l+v_l^{*}(t,x)\hat{a}_l^{out\,\dagger}),
    \label{eq: baseout}
\end{equation}
where we have defined new bosonic operators $\hat{a}_n^{out}$ corresponding to a new notion of particles. The connection between the two basis, is given by $v_n = \sqrt{2\omega_n}\sum_m (B_n^{m} u_m + A_n^{m} u_m^{*})$. This can be introduced in Eq. (\ref{eq: baseout}) to find
\begin{equation}
    \hat{a}_l^{out} = \sqrt{2\omega_l}\sum_m (B_l^{m} \hat{a}_m^{in}+ A_l^{m\,*} a_m^{in\, \dagger}) , 
    \label{eq: bogoliubov}
\end{equation}
which is known as a Bogoliubov transformation \cite{Dodonov, estadoDeVacio}. By using this relation among operators, we can find the number of particles of well-defined momentum $k_n$ in the region corresponding to $t>t_f$
\begin{equation}
    \langle\hat{N}_l\rangle = \sqrt{2\omega_l}\sum_m ((1+N_m^0)|A_l^m|^2+N_m^0|B_l^m|^2).
    \label{eq: NumPart}
\end{equation}
The result implies that the number of particles inside the cavity can vary with the movement of the walls, even yet for the case of an initial vacuum ($N_m^0 = 0$ for all $m$) particles can be created by the movement of the external walls. This is what is commonly known as Dynamical Casimir Effect \cite{Dodonov, crocceScalar}. 

As for the computation of the number of particles after having moved the walls, we need to know $A_l^m$ and $B_l^m$. To do this, we need to give a continuous description of the field during the time interval when the walls move. Hence, we introduce the expansion of the $u_m(t,x)$ functions in the instantaneous basis Eq. (\ref{eq: baseInstantanea}) into Eq. (\ref{eq: campoIn}). Further, we introduce that particular expression of $\hat{\phi}$ into the wave equation. By considering the fact that the instantaneous basis is orthonormal,  we find the following differential equation for the time dependent coefficients
\begin{flalign}
    &\ddot{Q}_l^{(m)}+k_l^2 Q_l^{(m)}=\nonumber\\
    &=-\sum_n \Big(g_{nl} \Big(2\lambda \dot{Q}_n^{(m)}+\dot{\lambda}Q_n^{(m)}+h_{nl}\lambda^2 Q_n^{(m)}\Big) \Big)
    \label{eq: diferencialQ}
\end{flalign}
where
\begin{gather}
    \lambda = \frac{\Delta \dot{L}(t)}{L}, \\
    g_{nl} = L(\partial_{\Delta L}\Phi_n,\Phi_l)
    \label{eq: gnl}
\end{gather}
and
\begin{equation}
    h_{nl} = L^2(\partial_{\Delta L}^2\Phi_n,\Phi_l).
\end{equation}
This set of infinite coupled differential equations does not admit exact analytical solutions and so it must be solved by either employing analytical approximations or numerical methods \cite{perturbaciones, MSA, russer}. We have used both of these approaches which we shall briefly discuss in the following.

\subsection{Multiple scale analysis}

We firstly consider some analytical approximations. In order to employ these methods, we assume that the walls perform small harmonic oscillations ($\epsilon \ll \min\{L_1,L_2\}/L$ in Eq. (\ref{eq: trayectoria})) and search for solutions of the form
\begin{equation}
    Q_n^{(m)} = A_l^m(\tau) e^{i\omega_l t} + B_l^m e^{-i \omega_l t},
\end{equation}
where we have introduced a second slower time scale $\tau = \epsilon t$ following the \textit{multiple scale analysis} (MSA) procedure \cite{MSA}. After introducing these expressions in Eq. (\ref{eq: diferencialQ}) and solving for the first order in $\epsilon$, we obtain a system of infinite coupled differential equations for $A_l^m(\tau)$ and $B_l^m(\tau)$ which reads
\begin{eqnarray}
    \frac{dA_l^{m}}{d\tau} &=& -\frac{ \eta_l}{2} \delta(\Omega-2\omega_l) B_l^{m} \nonumber\\&-& \frac{\Omega}{2\omega_l}\sum_{n\neq l} g_{nl}\big(\omega_n-\frac{\Omega}{2}\big)\delta(\Omega-(\omega_n+\omega_l))B_{n}^{m} \nonumber \\ &-& \frac{\Omega}{2\omega_l}\sum_{n\neq l} g_{nl} \Big(\big(\omega_n+\frac{\Omega}{2}\big) \delta(\Omega-(\omega_l-\omega_n)) \nonumber\\&+& \big(\omega_n-\frac{\Omega}{2}\big) \delta(\Omega-(\omega_n-\omega_l)\big)\Big) A_n^{m} \label{eq: alm mov}
\end{eqnarray}
and
\begin{eqnarray}
    \frac{dB_l^{m}}{d\tau} &=& -\frac{\eta_l}{2} \delta(\Omega-2\omega_l) A_l^{m} \nonumber\\&-& \frac{\Omega}{2\omega_l}\sum_{n\neq l} g_{nl}(\omega_n-\frac{\Omega}{2})\delta(\Omega-(\omega_n+\omega_l))A_{n}^{m} \nonumber \\ &-& \frac{\Omega}{2\omega_l}\sum_{n\neq l} g_{nl}\Big(\big(\omega_n+\frac{\Omega}{2}) \delta(\Omega-(\omega_l-\omega_n)) \nonumber\\&+& \big(\omega_n-\frac{\Omega}{2}\big) \delta(\Omega-(\omega_n-\omega_l)\big)\Big)B_n^{m}. \label{eq: blm mov}
\end{eqnarray}

Additionally, we consider  initial conditions $A_l^m(0) = 0$ and $B_l^m(0) = \delta_{l}^m/\sqrt{2\omega_l}$, so that the description of the field is continuous with the one given by Eq. (\ref{eq: campoIn}) \cite{crocceScalar}.

We can note that there are three conditions for the frequency of oscillations of the walls ($\Omega$ in Eq. (\ref{eq: trayectoria})) in Eqs. (\ref{eq: alm mov}) and (\ref{eq: blm mov})  that provide us with non-trivial solutions for $A_l^{(m)}$ and $B_l^{(m)}$. The first of these conditions is often known as single-mode resonance and reads
\begin{equation}
    \Omega = 2\omega_n. 
    \label{eq: Resonancia}
\end{equation}
Futher, there are two other conditions which are responsible of coupling different modes of the field, namely
\begin{align}
    \Omega &= \omega_m + \omega_n 
    \label{eq: suma}\\
    \Omega &= |\omega_m - \omega_n|.
    \label{eq: resta}
\end{align}

The first two of said conditions, Eqs. (\ref{eq: Resonancia}) and (\ref{eq: suma}), are responsible for the creation of particles. The third condition Eq. (\ref{eq: resta}), leads to the redistribution of photons between coupled modes but, by itself, it cannot lead to particle creation. Once $\Omega$ is set, more than just one of these conditions can be met. This is determined by the structure of the wave-number/eigenfrequency spectrum, and greatly affects both the rate at which the particles are created and the energy of said particles.

It is also important to note that the weight of terms associated with the single-mode resonance of mode $n$ is given by $\partial_{\Delta L}k_n$, which is to say, the slope of the wave number regarding  the displacement $\Delta L$. This quantity can be obtained from the transcendental equation (Eq. (\ref{eq: trascendente}))
\begin{align}
    \partial_{\Delta L} k_n = k_n \sin(k_n \Delta L)\Big( \big( L+\frac{2}{\alpha k_n^2}\big) \sin(k_n L) \nonumber\\- \Delta L \sin(k_n \Delta L)-\frac{2 L}{\alpha k_n}\cos(k_n L)\Big)^{-1}.
    \label{eq: derivkm}
\end{align}
On the other hand, the strength of the coupling between modes $n$ and $l$ is given, to a first order approximation, by $g_{nl}$ (Eq. (\ref{eq: gnl})).

\subsection{Numerical method}

We can generally solve Eq. (\ref{eq: diferencialQ}) by means of numerical methods. Herein,  we consider a method similar to the one used in \cite{numerical}, where some of us integrate the differential equations by employing a Runge-Kutta fourth order method between $t = 0$ and $t_{max}>t_f$. Hence, for $t>t_f$ the analytical solution to Eq. (\ref{eq: diferencialQ}) reads
\begin{equation}
    Q_n^{(m)}(t) = \frac{1}{\sqrt{2\omega_n}}(A_n^m(t_f)e^{-i\omega_n t}+B_n^m(t_f)e^{i\omega_n t}).
\end{equation}
Further we multiply by $e^{\pm i\omega_n t}$ and take the time-average in $t_f<t<t_{max}$, so as to obtain the values for $A_n^m(t_f)$ and $B_n^m(t_f)$. Introducing these values in Eq. (\ref{eq: NumPart}), we can then calculate the number of particles in the cavity at $t = t_f$, when the cavity is again at rest. Finally, by considering $t_f$ a continuous variable we can repeat this process for several time intervals and obtain $\langle N_n \rangle$ as a function of $t_f$.

It is important to recall that Eq. (\ref{eq: diferencialQ}) constitutes an infinite set of coupled differential equations. In order to perform the numerical integration of said equations we must consider a finite number of modes of the field, and neglect the rest. This in itself conforms an approximation, which, in order to be valid, must fulfill certain conditions. In particular, we consider the simulations to be valid until we see a considerable number of particles in a mode that fulfills any of the coupling conditions given by Eqs. (\ref{eq: suma}) and/or (\ref{eq: resta}) with a mode that is being neglected. The results presented in the following section have been obtained using the first 15 modes of the double cavity ($k_0$ to $k_{14}$), for both different initial configurations of the double cavity (meaning different values of $\alpha/L$ and $L_0/L$) and for different values of $\Omega$. We have set $\epsilon = 3 \times 10^{-3}$ as we have seen this value is consistent with the small oscillations condition $(\epsilon \ll \min\{L_1, L_2\}/L)$ for all values of $L_0$ considered.

\subsection{Results and discussion}

\begin{figure*}[t]
    \centering
    \includegraphics[width = \textwidth]{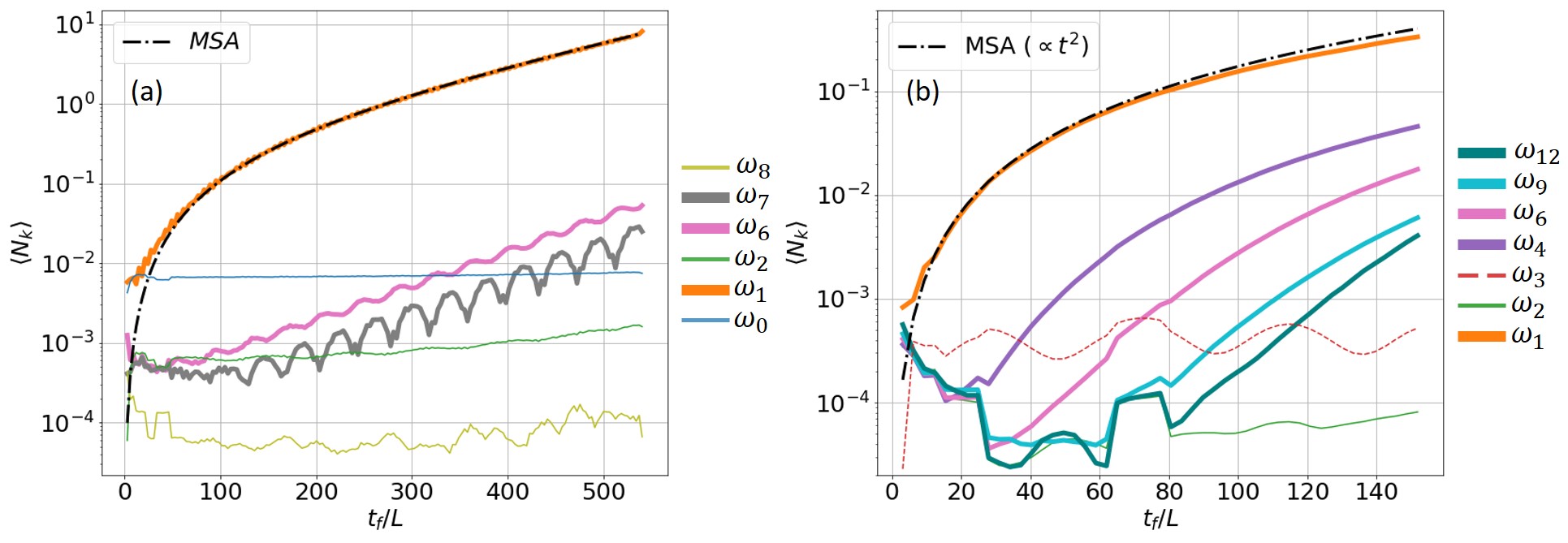}
    \caption{Number of particles as a function of the oscillation time ($t_f/L$) obtained through numerical simulations of the system considering an external frequency of $\Omega = 2\omega_1$ both for a non-equidistant spectrum (a) and an equidistant one (b). In (a), we can observe the results of the simulation for parameters: $\alpha/L = 0.5$ and $L_0/L = 0.44$. It is easy to note that we only get single mode resonance for $\omega_1$ and no other first order conditions. On the other hand, in (b) the double cavity considered has parameters $\alpha/L = 125$ and $L_0/L = 0.44$. In this case, we can consider all of the modes (with $n > 0$) to be localized in a single cavity, the solid lines representing modes that are localized in the left cavity, while the dotted ones stand for those that  are localized in the right one (exemplified here by mode 3, the behavior being analogous for the rest of them). Because of the equidistant spectrum in this second case, in addition to the single mode resonance for mode 1, we also get frequency subtraction coupling relations of the form $\Omega = \omega_{n+2}^{+}-\omega_n^{+}$ (as $\omega_{1}=\omega_{1}^{+}$) which allow for the creation of particles in modes 4, 6, 9 and 12. The rest of the modes corresponding to the left cavity present no particle creation and their behavior is exemplified here by mode 2. In both cases the analytical results for $\langle N_1 \rangle$ are shown by the black dot-dashed line. In (a) this was calculated by applying Eq. (\ref{eq: NumberRes}), while in (b) these results are valid only for short oscillation time ($t\ll 300$).}
    \label{fig: NumPartRes}
\end{figure*}

%Herein, we discuss the results obtained via the numerical simulations and contrast them with the analytical predictions. The results were obtained both for different initial configurations of the double cavity, meaning different values of $\alpha/L$ and $L_0/L$, and for different values of $\Omega$. We have set $\epsilon = 3 \times 10^{-3}$ as we have seen this value is consistent with the small oscillations condition $(\epsilon \ll \min\{L_1, L_2\}/L)$ for all values of $L_0$ considered.

We start with the case of a dielectric wall by setting $\alpha/L = 0.5$. For the lower modes this value of the susceptibility fulfills the condition $k_m \alpha \sim 1$. This results in a non-equidistant frequency spectrum for the less energetic modes. Firstly, we consider the external frequency to be satisfying the single-mode resonance condition for mode $1$ ($\Omega = 2\omega_1$). We choose mode 1 instead of the less energetic mode 0 because, as we have discussed, the latter vanishes for large enough values of the electric susceptibility. The field is considered to be in an initial vacuum state. The results of this numerical simulation can be seen in Fig. \ref{fig: NumPartRes} (a). Therein, we can observe the number of particles as a function of time for this value of $\alpha/L$ and $L_0/L = 0.44$. It is easy to note that there is an initially exponential growth in the number of particles only for the resonant mode. This feature agrees with the result predicted by solving Eqs. (\ref{eq: alm mov}) and (\ref{eq: blm mov}) for this case. Results are shown together with the numerical results as a black dot-dashed line, where the number of particles is predicted to grow as
\begin{equation}
    \langle N_1 \rangle = \sinh^2\Big(\frac{L\partial_{\Delta L}k_1(L_0)}{2}\epsilon t_f\Big).
    \label{eq: NumberRes}
\end{equation}
We note that this result is consistent with previous results for similar systems \cite{crocceScalar, russer, pau shaker}. The exponential growth of the resonant mode $\omega_1$ agrees with the result predicted by MSA, all other modes, do not exhibit exponential growth. However,  we can also see some particle creation process for other modes at later times ($t_f>300$). We can note that the number of photons in more energetic modes start growing, although they do so at a slower rate than for mode 1. The solutions obtained via the MSA do not predict photon number growth in modes other than the resonant one, and the result seen in Fig. \ref{fig: NumPartRes} corresponds to behaviors that can be described by higher order methods. In any case, for this choice of $\alpha/L$ and $L_0/L$ we manage to get exponential creation of photons from an initial vacuum state, which is the expected result for this choice of $\Omega$ and a non-equidistant spectrum.

Following, we can assume identical values for $L_0/L$ and $\Omega$, while considering a much higher value of the electric susceptibility, $\alpha/L = 125$. The results obtained for this new choice of susceptibility can be observed in Fig. \ref{fig: NumPartRes} (b). Apart from the zeroth mode, which is expected not to be localized, the solid line represents modes localized in the left cavity, while the dotted lines indicate those localized in the  right side. Additionally, we show the MSA prediction for $\langle N_1 \rangle$ which is valid for short time scales ($t_f \ll 300$). We can immediately notice a difference in the behavior of the system from the one observed for $\alpha/L = 0.5$ (Fig. \ref{fig: NumPartRes} (a)). It is easily seen that for a short time period ($\tau \ll (\epsilon \Omega)^{-1}$), the number of particles in mode $1$ grow following the MSA prediction, at a rate proportional to $t^2$. For longer time periods,  $\langle N_1 \rangle$ grows at a slower rate ($\propto t$), as well as those higher energy modes that fulfill the condition given by Eq. (\ref{eq: resta}).

For this value, all frequencies (other than $k_0$) can be considered to take their limit values given by Eq. \ref{eq: kmLoc} and hence, the spectrum in this case becomes equidistant. As this happens, we find that our choice of $\Omega$ no longer exclusively fulfills the single-mode resonance condition (Eq. (\ref{eq: Resonancia})), but also frequency subtraction coupling conditions (Eq. (\ref{eq: resta})) for infinite pairs of left cavity modes. This is a direct consequence of each of the left cavity's spectrum equidistance, as $\omega^{+}_{k+2}-\omega^{+}_{k} = 2\omega_1 = \Omega$. Although the right cavity is also equidistant, we do not have an equivalent relationship with $\Omega$, because $L_2$ is not a multiple of $L_1$. 

As we have noted before, the effects associated with the type of coupling relevant to this case, is the redistribution of energy between coupled modes, while the resonance condition is the one responsible of particle creation. The first modes are the ones that are firstly populated, starting with the resonant mode, while as time goes on, particles are created in more energetic modes. This is consistent with similar cases that have already been studied in the bibliography \cite{crocceScalar,crocceVect}. It is worthy noting that, for $\alpha = 0$, we cannot get particle creation via single-mode resonance because $\partial_{\Delta L} k_m = 0$ for any mode $m$ (see Eq. (\ref{eq: derivkm})), as the wave-numbers of the field become independent from the relative size of the individual cavities.

\begin{figure*}[t]
    \centering
    \includegraphics[width = \textwidth]{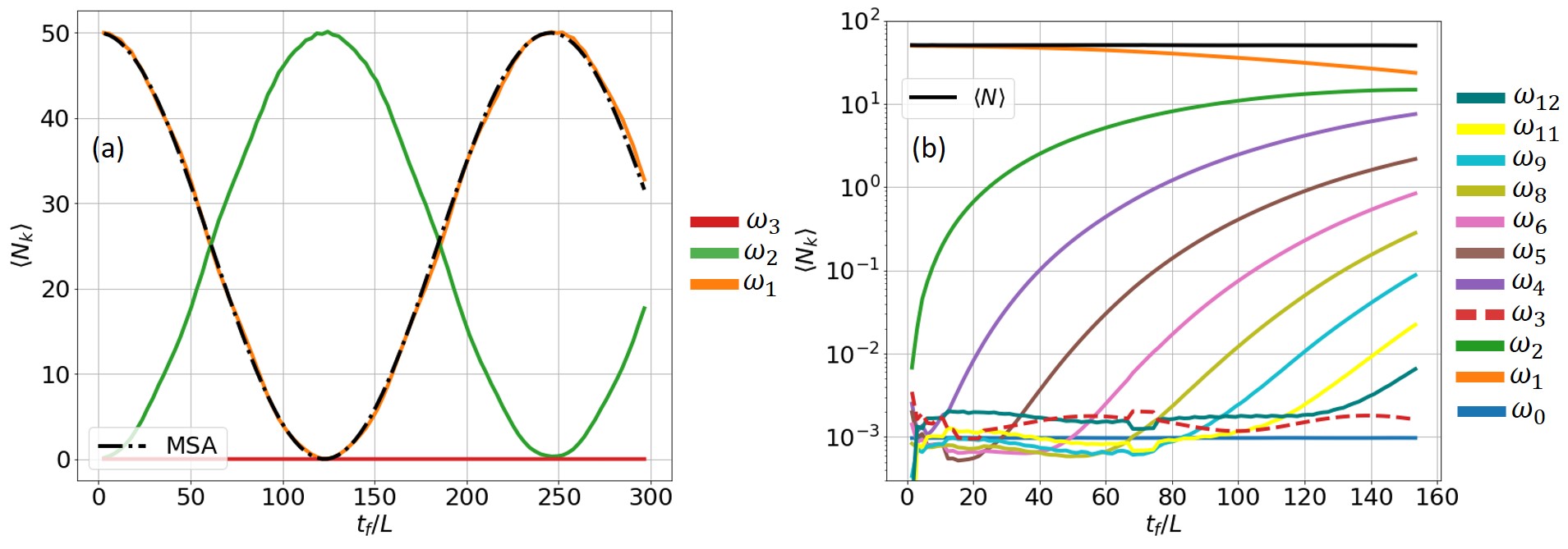}
    \caption{Number of particles as a function of oscillation time ($t_f/L$) obtained through numerical simulations of the system considering an external frequency of $\Omega = \omega_2 - \omega_1$ both for a non-equidistant spectrum (a) and an equidistant one (b). In this case the system starts with an initial number of particles in mode 1 $\langle N_1 \rangle(t_f = 0) = 50$. In (a) we see the results for $\alpha/L = 0.5$ and $L_0/L = 0.04$. In this case, only modes 1 and 2 couple which results in a harmonic migration of the photon number exclusively between said modes. The dashed-and-dotted black line stands for the MSA prediction for $\langle N_1 \rangle$ as seen in Eq. (\ref{eq: numpart restaMSA}). In (b) we again have the case where modes are localized in either the left (solid lines) or right (dashed lines) cavity  ($\alpha/L = 125$ and $L_0/L = 0.44$). As a consequence of the spectrum being equidistant in each cavity and as $\omega_{1,2} = \omega_{1,2}^{+}$  we get frequency subtraction coupling for every pair of successive modes in the left cavity ($\Omega=\omega_{n+1}^{+}-\omega_n^{+}$). This results in the migration of the particles that are initially in mode 1 to left-cavity higher energy modes as the wall oscillation time increases. The solid black line represents the total number of photons ($\langle N \rangle = \sum_k \langle N_k\rangle$) inside the double cavity which, as we can see, remains constant for all values of $t_f/L$ considered here.}
    \label{fig: NumPartResta}
\end{figure*}

We now consider the case where $\Omega = \omega_2 - \omega_1$. As we have previously mentioned, this type of coupling by itself will not lead to the creation of particles. Because of this, we consider that initially, mode 1 is populated by a certain number of photons ($N_1^0=\langle N_1 \rangle(t<0) = 50$). Again, we start by considering the case with the dielectric wall ($\alpha/L \sim 0.5$) and $L_0/L = 0.04$. This parameters will yield a non-equidistant spectrum. The results for the simulation of this case can be seen in Fig. \ref{fig: NumPartResta} (a). As we can see, in the absence of other coupling conditions, the photons oscillate harmonically between modes 1 and 2. This result agrees with the analytical results of Eqs. (\ref{eq: alm mov}) and (\ref{eq: blm mov}), which read
\begin{align}
    \expval{N_1} = N_1^0 \cos^2\Big(\frac{g_{21}(\omega_2^2-\omega_1^2)}{4\sqrt{\omega_1\omega_2}}\epsilon t_f\Big), \label{eq: numpart restaMSA} \\
    \expval{N_2} = N_1^0 \sin^2\Big(\frac{g_{21}(\omega_2^2-\omega_1^2)}{4\sqrt{\omega_1\omega_2}}\epsilon t_f\Big).
\end{align}
As we can see, the frequency at which the photons oscillate is proportional to the coupling coefficient between both modes, as can be expected. The numerical results also show no growth in $\langle N_k \rangle$ for any other mode which is consistent with the analytical results. 

On the other hand, we can see in Fig. \ref{fig: NumPartResta} (b) the results for the numerical simulation with the same choice of $\Omega$ but with $\alpha/L=125$ and $L_0/L=0.72$. In this case, all modes except for mode $0$ are localized in either the left (solid lines) or right (dashed lines) cavity. Now, for this particular choice of parameters, the spectrum of both individual cavities is equidistant. As modes 1 and 2 are localized both in the left cavity, for every pair of successive modes in the left cavity we get a coupling of the form $\omega_{n+1}^{+}-\omega_{n}^{+} = \Omega$. This results in the transference of photons from mode 1 to other modes with higher energy, starting with mode 2 ($\omega_2=\omega_2^+$), then going to mode 4 ($\omega_4=\omega_3^{+}$), etc. It must be noted that these results are also consistent with what we would expect the analytical solution to yield as the terms that contribute to it in Eqs. (\ref{eq: alm mov}) and (\ref{eq: blm mov}) will transfer the photons between the coupled modes while not creating any particles. This last fact is shown by the solid black line that stands for the total number of photons inside the double cavity (i.e. $\langle N \rangle = \sum_m \langle N_m \rangle$), which stays constant for all $t_f/L$.

Following, we can consider the case where $\Omega = \omega_1 + \omega_2$, fulfilling the coupling condition given by Eq. (\ref{eq: suma}). As we have previously mentioned, this results in the creation of photons at a rate that depends on the value coupling coefficient between these modes $g_{12}$ (Eq. (\ref{eq: gnl})). In the case of a non-equidistant spectrum, and in the absence of any other condition from Eqs. (\ref{eq: Resonancia}) or (\ref{eq: resta}), the number of photons is known to grow at the same exponential rate in both of the coupled modes as
\begin{equation}
    \langle N_1 \rangle = \langle N_2 \rangle = \sinh^2\Big(\frac{g_{21}(\omega_2^2-\omega_1^2)}{4\sqrt{\omega_1\omega_2}}\epsilon t\Big).
\end{equation}
This will be the case for values of $\alpha/L$ and $L_0/L$ for which $\alpha k_{1,2} \sim 1$. 

In the particular case of $\alpha = 0$, when the wall is completely transparent, because of the way the external walls move in phase, the system is commonly known as a \textit{shaker} \cite{pau shaker}. In this case, we get resonant coupling between modes $1$ and $2$, and subtraction coupling for $\omega_{k+5}-\omega_k$ which results in particle creation in an infinite number of modes for high enough times.

\begin{figure}[th!]
    \centering
    \includegraphics[width = 8cm]{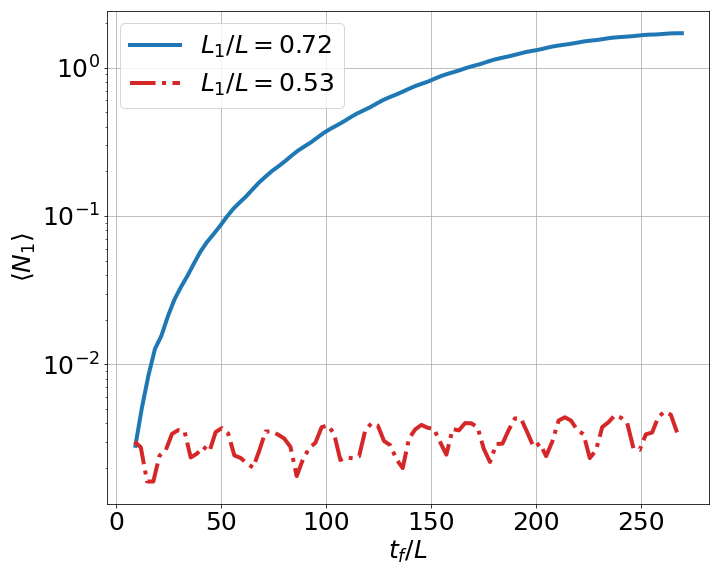}
    \caption{Number of particles of well defined frequency $\omega_1$ as a function of time for an electric susceptibility of $\alpha/L = 125$. The external excitation frequency is set at $\Omega = \omega_1+\omega_2$. The two values of $L_1/L$ correspond to, a case where modes 1 and 2 are localized in the same cavity ($L_1/L = 0.72$) and one where they are localized each in a different cavity ($L_1/L = 0.53$).}
    \label{fig: PärtNumSuma}
\end{figure}

For the opposite case of $\alpha/L \gg 1$, we might get different results depending on the value of $L_0$. In this case, the spectra of each of the individual cavities, becomes equidistant. Even more significant for particle creation, as the central wall resembles a perfect conductor, the two cavities are decoupled. This means that if modes 1 and 2 correspond to modes localized in different cavities, the coefficient $g_{12} = 0$, which  results in no particles being created. In Fig. \ref{fig: PärtNumSuma}, we can see $\langle N_1 \rangle$ for $\alpha/L = 125$ and two different values of $L_1/L$. It must be noted that although the photon number is shown only for mode 1, in the case of mode 2 the results are analogous.

One of these values, $L_1/L = 0.72$ corresponds to a particular configuration of the double cavity where modes $1$ and $2$ are localized in the same cavity. Here we can see $\langle N_1 \rangle$ grows with time. In this case, we also have a coupling with more energetic modes which fulfills $\omega_{k+3}^{+}-\omega_{k}^{+} = \Omega$, resulting in a similar situation to the one seen in Fig. \ref{fig: NumPartRes} (b). In the second case, $L_1/L = 0.53$, we do not see particle growth in the considered time interval. This is due to the fact that  modes 1 and 2 correspond to different cavities. We must note that the crucial difference among the two cases is which mode we are labeling as 2.  As the length of the left cavity increases (with the increase of $L_0$), the second mode of that cavity becomes less energetic than the first mode of the right cavity ($\omega_2^{+}<\omega_1^{-}$) and then, we label $\omega_2^{+} = \omega_2$ when, for lower $L_0$, we had $\omega_1^{-}=\omega_2$. 

As we have already mentioned, in the case where modes 1 and 2 correspond to different cavities we get $g_{12} = 0$ and thus, no particle creation. However, it might be  interesting, from an information point of view, to create particles in two cavities that are separated by a wall with $\alpha/L\gg 1$. This is because particles created via a frequency sum coupling condition are known to be created in entangled pairs. Hence, in principle,  achieving this would allow us to entangle two ''decoupled" cavities via the DCE. In the following subsection, we propose a way to do this by employing the residual zeroth mode in a cavity with a high value of $\alpha$.

Finally, we must note that this system allows for further studies in the field of particle creation. Some interesting examples would be to consider the movement of the dielectric wall while the perfectly conducting walls stay still, the consideration of a time-varying electric potential ($v(t)$) applied to the dielectric wall, or even the combination of these two (or other) time varying boundary conditions.

\subsection{The zeroth mode as a coupling tool} \label{sec: modo0}

As we have seen in Section \ref{sec: spectral}, in cases where $\alpha/L\gg 1$ we get a different behavior for modes with $m \geq 1$, tending to localize in either the left or right cavity. The behavior of the mode $m = 0$ is different since instead of localizing, it slowly vanishes as $(\alpha/L)^{-1/2}$. This exceptional feature allows us to consider a cavity whit a finite value susceptibility  but large enough for us to take $\omega_{m}^{\pm} = 2\pi/(L\pm L_0)$. This permits a description of  the double cavity as a system composed of two cavities which only 'share' mode 0, with the rest of the modes being localized in one or the other.

This scheme would allow to create particles in both cavities fundamental modes by indirectly coupling them via the zeroth mode. We may achieve said coupling by choosing a cavity configuration where
\begin{equation}
    \omega_1^{+}+\omega_0 = \omega_1^--\omega_0.
\end{equation}
The strength of the coupling between modes 0 and 1$\pm$ is mediated by $g_{01}^{\pm} = 1/(2\pi(\alpha(L\pm L_0))^{1/2})$ which is non-zero for finite values of $\alpha$. By using Eqs. (\ref{eq: kmLoc}) and (\ref{eq: modo0}), we find that this can be obtained by choosing 
\begin{equation}
    L_0/L = 1/\sqrt{(1+\pi^2\alpha/L )}.   
    \label{eq: L0Creacion0}
\end{equation}

\begin{figure}[th!]
    \centering
    \includegraphics[width = 8cm]{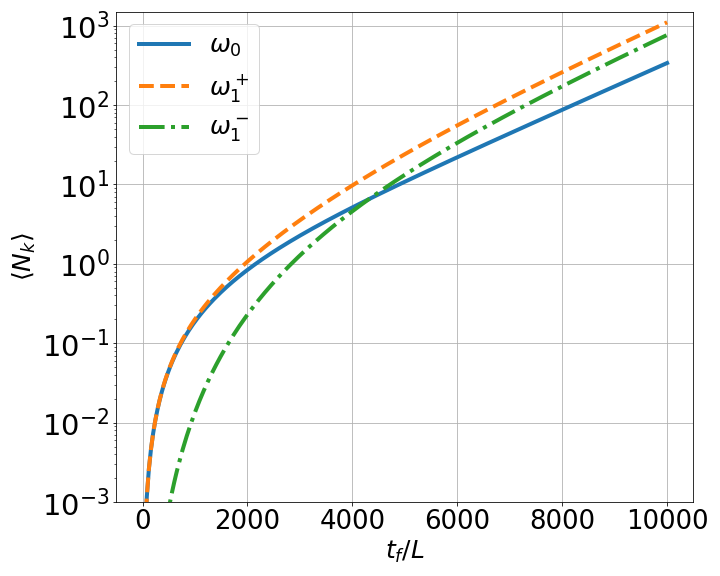}
    \caption{Number of particles as a function of oscillation time $t_f/L$ in the case where the external frequency is $\Omega = \omega_1^{+}+\omega_0$, the susceptibility is $\alpha/L = 30$ and $L_0/L$ corresponding to Eq. (\ref{eq: L0Creacion0}). The three modes plotted are those with non-trivial analytic solutions.}
    \label{fig: SumaModo0}
\end{figure}

The results for the number of particles can be obtained by employing the MSA method, and the value of $\langle N_m \rangle$ obtained doing this can be seen in Fig. \ref{fig: SumaModo0}. Here we can see that it is possible indeed to create particles in the different ``decoupled" cavities by employing the method above proposed. The number of photons would grow exponentially. However, the growth occurs at a very slow rate, and thus it requires to maintain the oscillations for long time intervals in order to get an appreciable number of particles.

Finally, we must note  that the zeroth mode could, in principle be used to overcome one of the practical limitations for the experimental detection of the mechanical DCE in a cavity. It is known that the values of $\Omega$ needed to create particles are usually higher than what current technology allows (the lowest being usually $\Omega = 2(\pi/L)$). In this case however, we can profit by the existence of the zeroth mode. For $\alpha/L = 30$ and $L_1/L = 0.74$, we get $\omega_0 = \omega_1^+/10$. In this case, we can choose $\Omega = 2\omega_0$ (which is an order of magnitude smaller than $2\omega_1^+$). Since we have $\partial_{\Delta L}k_0 = \Delta L k_0/(4L_1L_2) \neq 0$ we can obtain particle creation in mode 0 from single-mode resonance. There is, however, a persisting problem: the time for which the walls must be moving at this frequency. Through the MSA we can estimate that, even for times of $t/L \sim 2000$, we can obtain $\langle N_0 \rangle \sim 1$ for this choice of parameters. Going back to Fig. \ref{fig: NumPartRes} (a), we can observe  that we would get $\langle N_1 \rangle \sim 1$ for $t/L \sim 300$. This difference is due to the small value of $\partial_{\Delta L}k_0$ when $k_0$ takes small values and it implies that although oscillations can be maintained at a considerably smaller frequency, they would have to be maintained for larger time scales to get a measurable number of photons. 

\section{Entanglement} \label{sec: entanglement}

As we have mentioned, superconducting circuits have lately become extremely relevant in problems relating to quantum information and quantum communication, mostly because of their role in recent advances in quantum computing \cite{computacionChina, computacionCanada}. Particularly, techniques as boson sampling \cite{bosonSampling} present a promising alternative to achieve quantum supremacy. Recently, DCE has been studied in systems that employ boson sampling \cite{bosonSamplingDCE}. Considering all of this, we might note that the study of the relationship between the DCE and quantum information in this type of systems  might yield practical applications in the manipulation of entanglement.

We will start this section by providing the analytical tools to study the entanglement between pairs of modes of the field inside the double cavity. With this purpose we will employ gaussian state formalism. Gaussian states are those for which their characteristic functions and quasiprobability distributions are given by gaussian functions in phase space \cite{gaussianosteoria}. This subset of quantum states includes thermal, coherent, squeezed and vacuum states. 

From now on we will employ a basis of quadrature operators $R = (q_0,p_0,q_1.p_1,...)$, where
\begin{align}
    q_m = \frac{1}{\sqrt{2}}(a_m+a_m^{\dagger}) \\
    p_m = \frac{-i}{\sqrt{2}} (a_m-a_m^{\dagger}).
\end{align}
A given gaussian state $\rho$ can be completely characterized by its displacement vector \cite{gaussianosteoria}
\begin{equation}
    D_j = \expval{R_j}_\rho,
\end{equation}
and its covariance matrix
\begin{equation}
    V_{ij} = \frac{1}{2}\expval{R_i R_j+R_j R_i}_\rho - D_i D_j.
\end{equation}
The particular subset of gaussian states composed of those obtained through Bogoliubov transformations from an initial vacuum, we find that $D_j = 0$ for all $j$, so we shall assume this in the following. If we consider the field inside the double cavity as our global system, we can express the covariance matrix of the whole system in terms of $2\times 2$ matrices. The diagonal blocks correspond to the covariance matrix of a subsystem conformed only by a single mode $k$ of the field, this being
\begin{equation}
    V_k = \frac{1}{2}\begin{pmatrix} 2\expval{\hat{q}_k^2} & \expval{\hat{q}_k\hat{p}_k+\hat{p}_k\hat{q}_k} \\ \expval{\hat{q}_k\hat{p}_k+\hat{p}_k\hat{q}_k} & 2\expval{\hat{p}_k^2}\end{pmatrix}.
\end{equation}
The off-diagonal blocks can be though as the correlation between a pair of modes $j$ and $k$, which reads
\begin{equation}
    C_{jk} = \frac{1}{2}\begin{pmatrix} 2\expval{\hat{q}_k\hat{q}_j} & \expval{\hat{q}_j\hat{p}_k+\hat{p}_k\hat{q}_j} \\ \expval{\hat{q}_k\hat{p}_j+\hat{p}_j\hat{q}_k} & 2\expval{\hat{p}_k\hat{p}_j}\end{pmatrix},
\end{equation}
where we have $C_{kj} = C_{jk}^T$. 

If a system is in a global gaussian state, the states of subsystems are also gaussian and their covariance matrix is given by the restriction of the covariance matrix of the whole system to the subsystem that is being considered \cite{entNico}. 

In this section we start by considering the entanglement between two modes of the field $j$ and $k$. We thus consider a subsystem consisting of said pair of modes which is in a gaussian state characterized by
\begin{equation}
    V_{j|k} = \begin{pmatrix} V_j & C_{jk} \\ C_{jk}^T & V_k\end{pmatrix}.
\end{equation}
We call this type of subsystems a bipartite system. In order to quantify the entanglement between pairs of modes of the field, we employ the logarithmic negativity, which for bipartite systems can be defined as \cite{entNico}
\begin{equation}
    \mathcal{N}_{j,k} = \max\{0,-\ln(2\nu_{-})\}
    \label{eq: negativity}
\end{equation}
for this type of systems. In Eq. (\ref{eq: negativity}) we have defined
\begin{equation}
    \nu_{-} = \sqrt{\frac{\Sigma}{2}-\frac{1}{2}\sqrt{\Sigma^2-4\det V_{i|j}}}
\end{equation}
and
\begin{equation}
    \Sigma = \det V_j \det V_k - 2 \det C_{jk}.
\end{equation}

\subsection{Decoupling the systems}

In this section, we  study the problem of entanglement generation between the two halves of the double cavity by employing a protocol inspired by that presented in \cite{HalfEmpty}. We thus consider the double cavity we have been discussing in the previous sections of total length $L$, electric susceptibility $\alpha/L$, and the relative size of the single cavities given by $\Delta L(t<0) = L_0$. Initially ($t<0$, $in$) the cavity is stationary and the field is in a vacuum state. Starting at $t = 0$ and for an interval of time equal to $t_f$, the external perfectly conducing walls will move following a trajectory given by $\Delta L(0<t<t_f) = L_0 + L \epsilon \sin(\Omega t)$. At $t = t_f$ ($out$) the walls  stop at their initial position. At this same instant, we quickly turn on the potential difference applied on the central wall in such a way that $v\rightarrow\infty$. By looking at Eq. (\ref{eq: trascendente}) we can see that suddenly applying an very high potential difference at the wall has the same effect that taking $\alpha \rightarrow \infty$ if $v$ is high enough ($v/k_m \gg 1$). So, this last transformation is equivalent to turning the central dielectric wall into a perfect conductor, decoupling both cavities.

The description of the modes of the field after the movement of the external walls and before turning on $v$ is given in terms of the global eigenfunctions (Eq. (\ref{eq: NormalModes})). Once $v$ is turned on both halves of the cavity become decoupled, and so the eigenfunctions now become the localized ones, given by Eq. (\ref{eq: LocalizedModes}). The functions of both bases are related to each other by \cite{teoriaGlobalLocal}
\begin{align}
    u_l^\pm=\sum_m (\xi^{\pm}_{ml}u_m(t,x)+\chi^{\pm}_{ml}u_m^*(t,x)).
    \label{eq: LocVSGlob}
\end{align}
The expansion coefficients have been defined by using the Klein-Gordon product
\begin{align}
    \xi_{ml}^{\pm} = (u_l^{\pm},u_m)_{KG} \\ \chi_{ml}^{\pm} = -(u_l^{\pm},u_m^*)_{KG},
\end{align}
evaluating we get
\begin{align}
    \xi_{ml} = -e^{-i(\frac{\pi l}{L_{1/2}}-k_m)\Delta t}\Big(\frac{\pi l}{L_{1/2}}-k_m\Big)\mathcal{V}^{\pm}_{ml} \\
    \chi_{ml} = e^{-i(\frac{\pi l}{L_{1/2}}+k_m)\Delta t}\Big(\frac{\pi l}{L_{1/2}}+k_m\Big)\mathcal{V}^{\pm}_{ml} 
\end{align}
where $\Delta t = t-t_f$ and 
\begin{equation}
    \mathcal{V}_{ml}^{\pm} = (-1)^{l}\frac{\sin(k_m L_1)\sin(k_m L_2)}{\sqrt{2 L_{1,2}N_m ((\frac{\pi l}{L_{1,2}})^2-k_m^2)}}\sqrt{\frac{\pi l}{L_{1,2} k_m}}
\end{equation}
if $k_m \neq \pi l /L_1$ and $k_m \neq \pi l/L_2$, while if $k_m = \pi l/L_1$
\begin{equation}
    \mathcal{V}_{ml}^+ = \frac{L_1}{2\pi N_m}\sqrt{\frac{L_1}{2}}\sin\Big(\frac{L_2}{L_1} l\pi\Big),
\end{equation}
and if $k_m=\pi l/L_2$
\begin{equation}
    \mathcal{V}_{ml}^- = \frac{L_2}{2\pi N_m}\sqrt{\frac{L_2}{2}}\sin\Big(\frac{L_1}{L_2} l\pi\Big).
\end{equation}
Both basis of eigenfunctions are associated to their own bosonic operators, which correspond to the notion of particle before and after we turn $v$ on. Said operators can be associated via a Bogoliubov transformation which employ coefficients $\xi_{ml}^{\pm}$ and $\chi_{ml}^{\pm}$
\begin{equation}
    \hat{a}_{l\pm} = \sum_m (\xi_{ml}^{\pm*} \hat{a}_m^{out} - \chi_{ml}^{\pm *} \hat{a}_m^{out\, \dagger}).
    \label{eq: bogLoc}
\end{equation}
The bosonic opperators $a^{out}$ correspond to the global eigenfunctions for $t=t_f$ and can be readily linked to those corresponding to the initial state of the system ($a_m^{in}$) by Eq. (\ref{eq: bogoliubov}). We note that the bogoliubov trasnformation of Eq. (\ref{eq: bogLoc}), hints at the creation of particles because of taking $v\rightarrow\infty$ at $t = t_f$. Indeed, by substituting $\sqrt{\omega_l}A_l^m$ and $\sqrt{\omega_l}B_l^m$ by $\xi^{\pm}_{ml}$ and $\chi^{\pm}_{ml}$ respectively in Eq. (\ref{eq: NumPart}) we get
\begin{align}
    \langle N_{l\pm} \rangle= \frac{4\pi}{(L\pm L_0)}\sum_m \Big[N_m^0\Big(\frac{2\pi l}{(L\pm L_0)}-k_m \Big)^2\\+(1+N_m^0)\Big(\frac{2\pi l}{(L\pm L_0)}+k_m \Big)^2\Big] \mathcal{V}_{ml}^{\pm \, 2}.
\end{align}
The result above implies that even if no photons are created during the oscillation of the walls ($0<t<t_f$), we will nevertheless get particle creation as a result of suddenly turning on $v$. In reference \cite{HalfEmpty} it was shown that, in the case of an initial single cavity $\alpha = 0$ entanglement can be harvested from the vacuum state.

By replacing $a_m^{out}$ with Eq. (\ref{eq: bogoliubov}) in Eq. (\ref{eq: bogLoc}), we can relate the local bosonic operators with the ones corresponding to the initial state of the system, linking the two transformations that result from moving the walls and turning $v$ on. This results in
\begin{equation}
    a_{j\pm} = \sum_m [\tilde{\alpha}^{\pm}_{jm} \hat{a}_m^{in}+\tilde{\beta}^{\pm*}_{jm} \hat{a}_m^{in\, \dagger}]
\end{equation}
where we have defined
\begin{align}
    \tilde{\alpha}_{jm}^{\pm} = \sum_l \sqrt{2\omega_l}(B_l^m\xi_{lj}^{\pm *}-A_l^m\chi_{lj}^{\pm *}) \\
    \tilde{\beta}_{jm}^{\pm *} = \sum_l \sqrt{2\omega_l}(B_l^m\xi_{lj}^{\pm *}-A_l^m\chi_{lj}^{\pm *}).
    \label{eq: bogoliubovFull}
\end{align}

Since we aim to study the entanglement between modes localized in different cavities, in the following we shall consider bipartite systems formed by one mode $j+$ in the left cavity and another one $k-$ in the right cavity. Taking this into account, and as a way to simplify notation when calculating the logarithmic negativity $\mathcal{N}_{j,k}$ we will consider the first index to always correspond to a mode of the left cavity, while the other one will correspond to the one in the right.

\subsection{Results and discussion}

The results showcased in this section where obtained by numerically evaluating the expressions presented before. To do this we must consider a finite number of modes, in this case taken as $N = 200$. We must note that, in the same way as with the numerical simulations of Section \ref{sec: creation} taking a finite $N$ represents an approximation. This approximation can be justified by considering that the potential difference $v$ cannot be turned on in an infinitely short time. In practice, this transformation is done in a certain finite time interval $\delta t$. However, we can consider that for mode $j$ the transformation happens in a negligible amount of time if $\delta t \ll \omega_j^{-1}$ \cite{HalfEmpty}. This means that while for less energetic modes the transformation is instantaneous, for those with higher energies the transformation will be slower. Indeed, for the more energetic ones, the transformation (of turning on the potential difference) can be considered to be adiabatic and as such it will neither create particles nor generate entanglement. 

Before going on, we also note that the transformation of instantaneously introducing a potential difference in the position of the wall can be easily translated into circuits. In this last case, said transformation would consist in turning $E^{0\prime}(t)$ from $0$ to a value much higher than $1$. This would indeed be possible as it is similar to what is used to simulate the perfectly conducting walls at the extremes of the circuit \cite{squid1,squid2}.

\begin{figure}[th!]
    \centering
    \includegraphics[width = 8cm]{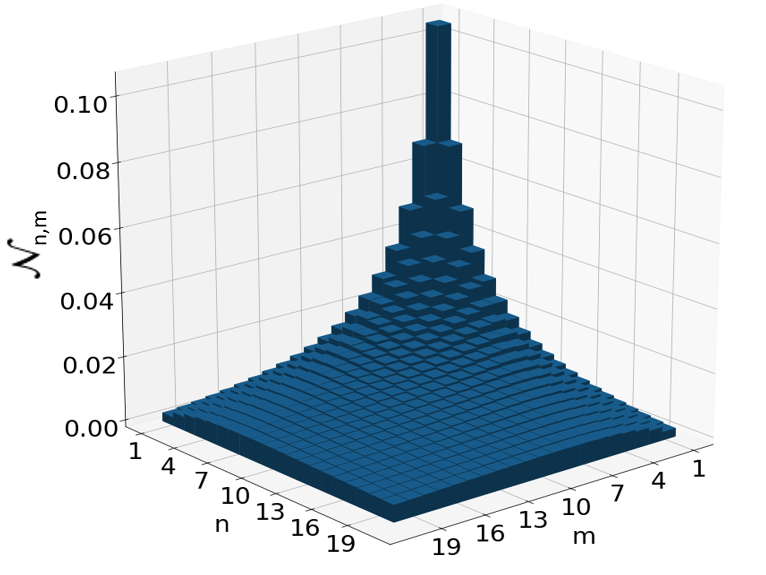}
    \caption{Logarithmic negativity for bipartite systems formed by pairs of modes localized in different cavities for an initial double cavity with $L_0/L = 0$ and $\alpha/L = 0$. Only the first 20 modes corresponding to each cavity are shown.}
    \label{fig: EntCasoControl}
\end{figure}

We start by considering the case where the perfectly conducting walls of the double cavity remain stationary at all times. In this case, the only transformation that the system will undergo is the turning on of $v$ at $t = 0$. We start by considering the simplest case, where the double cavity is in a symmetrical configuration ($L_0 = 0$) and the dielectric wall is fully transparent ($\alpha/L = 0$). In Fig. \ref{fig: EntCasoControl} we present the values of the logarithmic negativity for the first 20 modes localized in the two cavities. This case was presented in \cite{HalfEmpty} and works as a control case in relation for the variations we will consider later.

As we can see, the entanglement structure in this case is symmetrical, with the larger values of $\mathcal{N}_{n,m}$ concentrated among pairs of less energetic modes. Moreover, we see that the maximum of the entanglement corresponds to the one between the fundamental modes of each cavity ($m = n = 1$). As the difference between $n$ and $m$ increases the entanglement between modes tends to vanish at a slow rate. 

\begin{figure}[th!]
    \centering
    \includegraphics[width = 8cm]{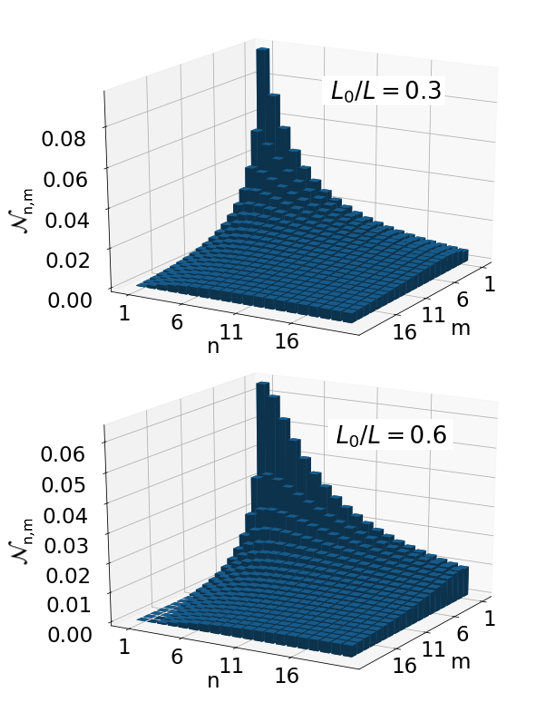}
    \caption{Logarithmic negativity for bipartite systems formed by pairs of modes localized in different cavities for an initial double cavity with $\alpha/L = 0$ and different values of $L_0/L$. Only the first 20 modes corresponding to each cavity are shown.}
    \label{fig: EntCasoDL}
\end{figure}

In Fig. \ref{fig: EntCasoDL} we show $\mathcal{N}_{n,m}$ for two different initial double cavity configurations with different values for $L_0/L$. As we can see, by increasing the value of $L_0$ and as a result, making the two cavities differ in size, the entanglement looses the symmetrical structure it had in Fig. \ref{fig: EntCasoControl}. This happens because, as $L_0/L$ increases so does the length of the left cavity, $L_1$, which results on the frequencies corresponding to the modes of this cavity being more similar to each other as $\omega_{n+1}^+-\omega_n^+\propto L_1^{-1}$. The opposite happens with the right cavity, which gets smaller and so the frequencies of its corresponding modes become more different among each other $\omega_{m+1}^{-}-\omega_m^-\propto L_2^{-1}$. This  results in $\mathcal{N}_{n,m}$ a slower variation with $n$, and a faster one with $m$. In the case for $L_0 = 0.6$, we can see that the negativity vanishes for certain pairs of modes, for example $n = 1$ and $m = 12$. Apart from this, we can see in both cases shown that the maximum is still located in $n=m=1$, although its value is lower than for the symmetric case from $0.1$ to $0.06$.

\begin{figure}[th!]
    \centering
    \includegraphics[width = 8cm]{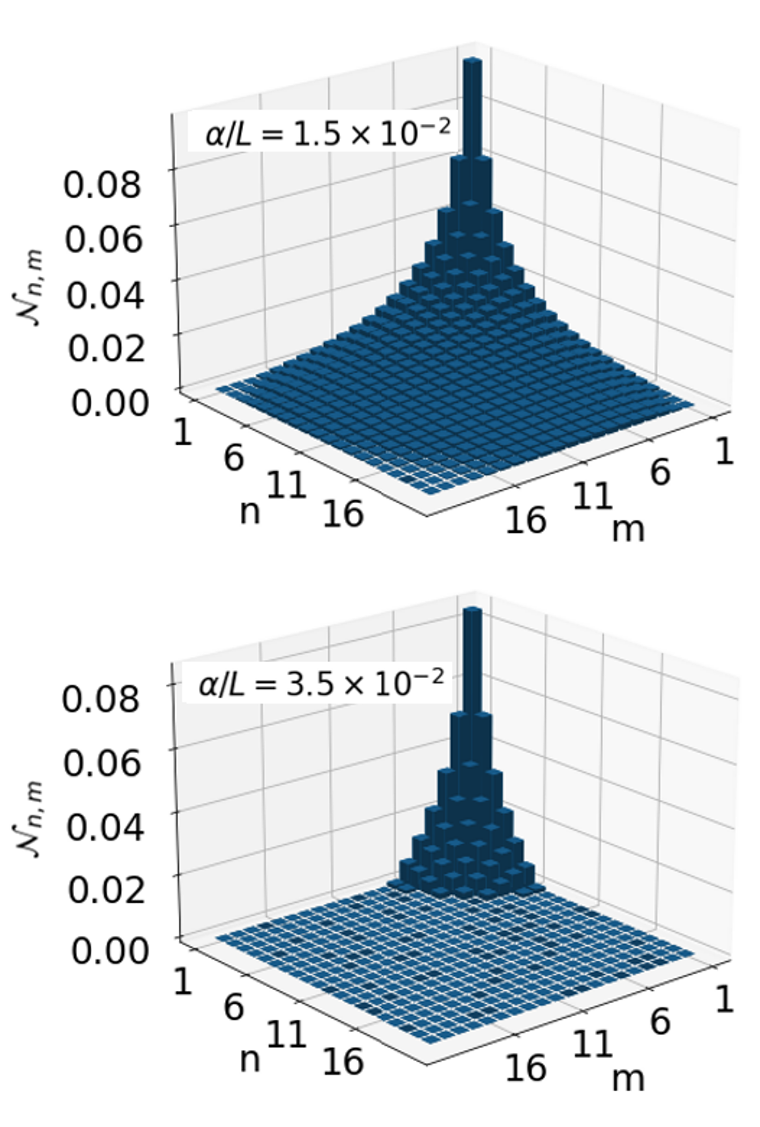}
    \caption{Logarithmic negativity for bipartite systems formed by pairs of modes localized in different cavities for an initial double cavity with $L_0/L = 0$ and different values of $\alpha/L$. Only the first 20 modes corresponding to each cavity are shown.}
    \label{fig: EntCasoAlfa}
\end{figure}

We further consider the case where initially, $L_0/L = 0$ but, $\alpha/L \neq 0$. The results are shown in Fig. \ref{fig: EntCasoAlfa}. We can see the inclusion of a non-transparent wall in the initial system results in a decrease in the value of $\mathcal{N}_{n,m}$ for all bipartite systems, lowering the maximum in $n = m = 1$ from $0.1$ to $0.08$. In fact, for many pair of modes the entanglement vanishes as we increase $\alpha/L$. This can be understood if we consider that, as we increase the susceptibility of the central wall, the initial configuration becomes more similar to the final system where the central wall is essentially a perfect conductor. In fact, as we have previously discussed, for higher modes, the values of $\alpha/L$ required to get their eigenfunctions localized are lower. As such, for these modes the transformation of turning on $v$ is essentially an adiabatic transformation as neither their eigenfunctions nor eigenfrequencies are affected. This results in no particles being created in those modes and thus no entanglement with the modes located in the other cavity. Finally, we note that in these case the symmetry of the entanglement structure is maintained.

Having studied how the initial state of the cavity modifies the entanglement structure we obtain after separating the two halves, we might now  consider the case where the external walls of the cavity move for a certain amount of time. We assume that the initial configuration of the cavity is given by the simplest case of $L_0/L = 0$ and $\alpha/L = 0$. We consider the walls oscillate at a frequency given by $\Omega = \omega_1+\omega_2 = 5\pi/L$. As we mentioned before, after the oscillation we turn on $v$ decoupling the cavities. So as to keep this case as simple as possible, we assume a toy model of the field inside the cavity, for which we dismiss the frequency subtraction couplings that this choice of $\Omega$ would satisfy. This means the energy delivered to the field by the movement of the walls only results in particle creation in modes 1 and 2. It is important to  note that although this is not possible in a one-dimensional cavity it might be in a three-dimensional one.

\begin{figure}[th!]
    \centering
    \includegraphics[width = 9cm]{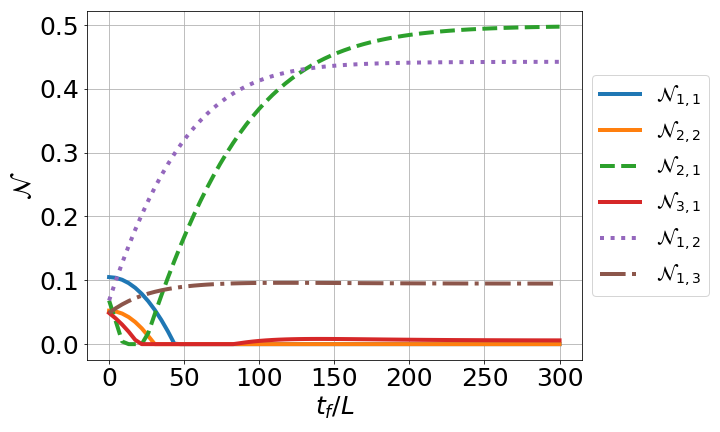}
    \caption{Logarithmic negativity for bipartite systems formed by pairs of modes localized in different cavities for an initial double cavity with $L_0/L=0$ and $\alpha/L = 0$, where at $t=0$ the external perfectly conducting walls oscillate at a frequency $\Omega = \omega_1+\omega_2$ until $t=t_f$.}
    \label{fig: EntDCE}
\end{figure}

The non-trivial Bogoliubov coefficients associated to the particle creation read
\begin{align}
    \sqrt{2\omega_1}A_1^2 = \sqrt{2\omega_2}A_2^1= \sinh\Big(\frac{|\gamma_{12}|}{2}\tau\Big)  \\
    \sqrt{2\omega_1}B_1^1 = \sqrt{2\omega_2}B_2^2= \cosh\Big(\frac{|\gamma_{12}|}{2}\tau\Big).
\end{align}
By introducing the above in equation \ref{eq: bogoliubovFull} we can calculate $\mathcal{N}_{n,m}$ through the covariance matrix. In Fig. \ref{fig: EntDCE} we observe the results of $\mathcal{N}_{n,m}$ for this case. We can see that for large enough $t_f/L$, all the bipartite systems reach an asymptotic value of $\mathcal{N}_{n,m}$. For most subsystems, we get a vanishing value of $\mathcal{N}_{n,m}$. Although, in Fig. \ref{fig: EntDCE} only the cases with $n=m=1,2$ vanish, all of the pairs of modes that have been omitted show the same behavior for large enough values of $t_f/L$. For the other bipartite systems shown, we see that the negativity takes on non-zero asymptotic values. The maximum of said values is achieved for $n = 2$ and $m = 1$ ($\mathcal{N}_{2,1} \approx 0.49 $), followed closely by $n = 1$ and $m = 2$ ($\mathcal{N}_{1,2} \approx 0.44$), for the rest of the subsystems we get comparatively negligible values of negativity. Said values decrease with the difference between $n$ and $m$. 

It should also be noted that while $\mathcal{N}_{1,2}$ grows monotonically, $\mathcal{N}_{2,1}$ decreases at first, vanishing for a finite time interval, increasing again for larger values of $t_f/L$. This last fact, together with the fact that $\mathcal{N}_{1,2}\neq \mathcal{N}_{2,1}$ for $t_f/L\gg 1$, show an asymmetry in the entanglement structure, even if the final state of the system is conformed of two identical cavities. Interestingly, the asymmetry is dependent upon the initial velocity of the walls. It can be seen that if we take $\epsilon \rightarrow -\epsilon$ in Eq. (\ref{eq: trayectoria}), meaning that the walls move initially towards the right instead of towards the left ($L_{1,2}=(L\pm\Delta L)/2$), the roles of $m$ and $n$ in $\mathcal{N}_{n,m}$ are inverted. 

We can note that the maximum values taken by $\mathcal{N}_{1,2}$ and $\mathcal{N}_{2,1}$ are many times larger than their initial values. This shows that the DCE can be used to increase the entanglement between pairs of modes when compared to the case where there is no particle creation. We also find a simplification of the entanglement structure, as only two bipartite systems ($n=1$, $m=2$ and $n=2$, $m=1$) present a non vanishing logarithmic negativity.

\begin{figure}[th!]
    \centering
    \includegraphics[width = 8.5cm]{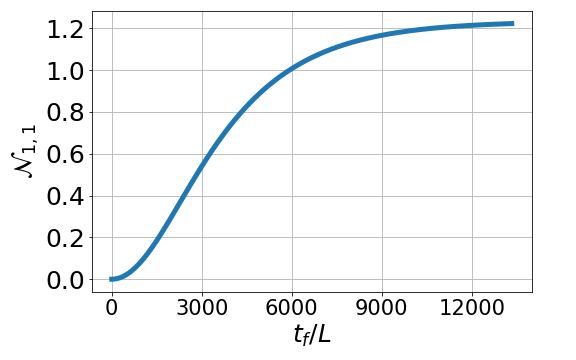}
    \caption{Logarithmic negativity for bipartite systems formed by pairs of modes localized in different cavities for an initial double cavity with $L_0/L=0$ and $\alpha/L = 30$, where at $t=0$ the external perfectly conducting walls oscillate at a frequency $\Omega = \omega_1^{+}+\omega_0=\omega_1^{-}-\omega_0$ until $t=t_f$.}
    \label{fig: EntCasoModo0}
\end{figure}

Another approach to entangle modes in different cavities is to use the method discussed in Section \ref{sec: modo0}. We have shown that by coupling the 1st localized modes in each cavity with the zeroth mode, it was possible  to get particle creation in modes that are localized in both cavities. The fact that the modes are initially localized requires for $\alpha/L$ to be high in the initial configuration of the double cavity. This means that turning on $v$ will not affect any mode besides mode zero, which must vanish when the walls are perfect conductors. In this case, we might consider that we take $v\rightarrow \infty$ slowly so as to neglect the entanglement that this non-localized mode might contribute. Because of this, it is easy to see $\mathcal{N}_{1,1}$ will be the only quantity that is not zero. In Fig. \ref{fig: EntCasoModo0} we show $\mathcal{N}_{1,1}$ as a function of $t_f$. 

In this case, we see a generation of entanglement between the two modes which are initially uncorrelated. As the oscillation time increases we find  that the negativity takes, for sufficiently large $t_f/L$, an asymptotic value of $\mathcal{N}_{1,1} \approx 1.23$. Said value is approximately $2.5$ times larger than the maximum obtained for the previous case ($\mathcal{N}_{2,1} \approx 0.49$). So, this approach results in stronger entanglement between localized modes if we consider a single bipartite system. Here, the structure is further simplified, as the only entangled subsystem is the one with $n = 1$, $m = 1$. However, the problem with this case compared with the previous one is the large values of $t_f/L$ needed to get comparatively higher values of $\mathcal{N}$. To achieve $\mathcal{N}_{1,1} = 0.5$, one would need $t_f/L \approx 3000$ which is at least 10 times greater than that needed to achieve the asymptotic value of $\mathcal{N}_{2,1} \approx 0.49$.

\section{Conclusions}\label{sec: conc}

Throughout this work we have studied the DCE in a double cavity, its link with quantum superconducting circuits and the possibility of using this effect in quantum information. We have exhibited the variability of the frequency spectrum, showing its dependence on the electric susceptibility $\alpha$ and on the relative size of the individual cavities ($L_1/L$ and $L_2/L$). We have shown that for extreme values of $\alpha$ analytical solutions exist for the eigenfrequencies. We have also obtained the spectrum numerically, showing the continuous behavior between the two extreme cases of the susceptibility. Furthermore, we have shown that for high enough values of $\alpha/L$ the modes localize in either the left or right cavity. This is true except for the least energetic mode that tends to vanish for high enough values of said parameter, but at a slower rate than the rest. Further, we have arrived at the conclusion that depending on the choice of parameters we can obtain both an equidistant structure ($\alpha/L \gg 1$ and $\alpha/L \ll 1$) as well as a non-equidistant one ($\alpha/L \sim 1$), which results in a very advantageous characteristic of the system.

Further, we have studied the  particle production via the DCE. This study 
has been developed both by analytical approximations (MSA) and exact numerical integration of the dynamical equations of the system Eq. (\ref{eq: diferencialQ}). We have discussed the three conditions over the external frequency $\Omega$ that can lead to non-trivial solutions, two of which lead to particle creation ($\Omega = 2\omega_m$ and $\Omega = \omega_m + \omega_n$) and one that only provides photon redistribution ($\Omega = |\omega_m - \omega_n|$). We have shown that changes in the structure of the spectrum lead to widely different results. For example, we have seen that in the non-equidistant case we can get exponential photon production through single mode resonance in the resonant mode, at a rate that is proportional to the slope of the wave-number in relation to changes in wall displacement ($\partial_{\Delta L}k_l$). In the equidistant case, we get photon production in the resonant mode that goes as a power law in time ($t_f^2$ for short times and $t_f$ for long times), but we also get particle creation in other, more energetic modes, as a result of coupling by subtraction condition. Beyond the analytical expressions of the number of particles as a function of $t_f$, the equations that resulted from this analysis can be used to predict and understand the behavior of the field. We have also proposed the study of other time-varying boundary conditions, such as a time-varying potential applied to the dielectric wall. We have considered the cases in which the zeroth mode can be employed as a coupling tool between both halves of the cavities, that are essentially decoupled for large susceptibilities. Furthermore, we have briefly discussed a way to excite the exponential creation of particles in this mode through single-mode resonance. It is important to stress that this  takes place for values of $\Omega$ which are one order smaller than what is usually considered for experimental proposals. However, whenever the zeroth mode is involved, much longer excitation times are needed to get a non-negligible number of particles. 

Finally, we have analyzed a protocol with which we can induce particle creation and later decouple the two cavities that conform our system. This protocol allows  to generate a system of two decoupled but entangled cavities, from an initial double cavity with a semi-transparent wall in the middle. We have shown how the entanglement gathered from an initial vacuum (in the absence of particle creation) can be modified by changes in the configuration of the double cavity, i.e. different values of $\alpha/L$ or $L_0/L$. Additionally, 
we have shown how the presence of DCE particles produced from the coupling by resonance condition ($\Omega = \omega_1 + \omega_2$) can be used to increase the entanglement between certain pairs of modes. For example, $\mathcal{N}_{1,2}$ increases from $0.06$ to $0.44$. We have also analyzed an alternative protocol, where we use the zeroth mode as an indirect coupling tool between modes in different cavities. In such a case, we have seen that entanglement can be generated between the modes that are localized in the two cavities and its maximum can be higher than the one obtained for the other protocol ($\mathcal{N}_{1,1}= 1.2$ in this case, and $\mathcal{N}_{1,1}= 0.49$ in the other). However, as in the case with particle creation involving the zeroth mode, the excitation times needed to improve the entanglement between the different cavity modes are much larger (2 orders of magnitude) than in the resonant case.

\section*{Acknowledgments}
This research was supported by Agencia Nacional de Promoción Científica y Tecnológica (ANPCyT), Consejo Nacional de Investigaciones Científicas y Técnicas (CONICET), Universidad de Buenos Aires (UBA). P.I.V. thanks The Abdus Salam International Center for Theoretical Physics for its support through the Associate programme. The research of F.C.L. was supported in part by the National Science Foundation under Grant No. PHY-1748958. We thank D. Blanco, E.A. Calzetta, and F.D. Mazzitelli for useful comments. %%%%%%%%%%%%%%%%%%%%%%%%%%%%%%%%%%%%%%%%%%%%%%%%%%%%%%%%%%%%%%%%%%%%%%%%%%
%%%%%%%%%%%%%%%%%%%%%%%%%%%%%%%%%%%%%%%%%%%%%%%%%%%%%%%%%%%%%%%%%%%%%%%%%%
%%%%%%%%%%%%%%%%%%%%%%%%%%%%% Appendix %%%%%%%%%%%%%%%%%%%%%%%%%%%
%%%%%%%%%%%%%%%%%%%%%%%%%%%%%%%%%%%%%%%%%%%%%%%%%%%%%%%%%%%%%%%%%%%%%%%%%%
%%%%%%%%%%%%%%%%%%%%%%%%%%%%%%%%%%%%%%%%%%%%%%%%%%%%%%%%%%%%%%%%%%%%%%%%%%

\end{document}